# Machine Learning Bandgap Prediction of Nanoporous Graphenes with Water

Sneha Mittal, *, † Alan E. Anaya Morales, †, ‡ Victor Rosendal, †, ‡ Mads Brandbyge *, †

†Department of Physics, Technical University of Denmark, 2800 Kgs., Lyngby, Denmark
‡These authors contributed equally to this work.
*Corresponding Author. E-mail: snemi@dtu.dk (S.M.); mabr@dtu.dk (M.B.)

**ABSTRACT**

The structure and dynamical behavior of water confined at or within nanostructures is a topic central to many fields from biology to emerging electronics such as carbon nanostructures. Nanoporous graphene (NPG) containing periodic nanoscale pores with specific topologies has emerged as a promising material in carbon-based nanoelectronics; however, its interaction with ambient water remains poorly understood. Here, we combine density functional theory (DFT), *ab-initio* molecular dynamics (AIMD), and interpretable machine learning (ML) to reveal how water controls quantum transport in NPGs. Depending on the local hydration structure, the bandgap varies by more than a factor of two across NPG and nitrogen-doped hybrid (h-NPG) systems. To uncover the underlying mechanism, we develop Smooth Overlap of Atomic Positions (SOAP)-based black-box and physics-informed grey-box ML models. The Gaussian process regression model achieves near-DFT accuracy ($R^2 \sim$ 0.94-1.00, RMSE $\lesssim$ 0.02 eV) while enabling physical interpretation. Analysis identifies water dipole orientation, water-substrate distance, water center-of-geometry, and ribbon-resolved dipole moments as the dominant factors controlling bandgap modulation across NPG and h-NPG systems.

## INTRODUCTION

At the nanoscale, the interaction of water with two-dimensional materials is crucial for applications such as ferroelectricity,[1,2] energy storage,[3] nanofluidics,[4] and water splitting.[5] Beyond these applications, interfacial and confined water can modulate electronic properties by altering local electrostatic potentials, charge redistribution, and molecular dipole ordering.[6–8] Recent experiments have demonstrated that nanoconfined water can undergo liquid-solid transitions even at ambient conditions, highlighting the profound impact of confinement on its structural and dynamical properties.[9] Furthermore, collective ordering and dynamics of water molecules have been shown to induce ferroelectric behavior in graphene nanoribbon devices.[1,2,10] This interplay between water and low-dimensional carbon systems becomes even more intriguing in nanoporous graphenes (NPGs), which feature dense, periodic arrays of nanoscale pores within single- or few-layer graphene.

NPGs enable tuning of electronic and transport properties, positioning them at the forefront of next-generation carbon nanoelectronics.[11–17] Unlike graphene, which lacks an intrinsic bandgap, NPG allows bandgap opening through structural design of the porous lattice, with electronic properties tunable via pore size, periodicity, edge chemistry, and the extent of $\pi$ delocalization in the framework.[18,19] In particular, the topology and chemical structure of the constituent graphene nanoribbons, together with in-plane anisotropy and quantum confinement effects, play a central role in determining the band structure and electronic transport characteristics.[20–23] Given the presence of water under realistic experimental conditions, understanding the interaction of water with NPGs is relevant.[24–27] Key questions include: How do spatial arrangement, density, position, and orientation of water molecules along the NPGs layer impact the band structures and electronic states? What is the underlying mechanism by which water modulates electronic transport? Do different types of NPGs exhibit similar sensitivity to water adsorption?

Addressing these questions requires a detailed understanding of NPGs in aqueous environments at the molecular level, for which advanced automated tools like machine learning (ML) can be helpful.[28–33] For 2D materials, it has been widely used to predict electronic, mechanical, and thermodynamic properties, including bandgaps, charge transport, and response to external stimuli such as doping, strain, or electric fields.[34–36] Herein, we develop an interpretable ML framework in combination with DFT and AIMD simulations to investigate how water controls electronic structure and quantum transport in NPGs. Our results reveal that the electronic structure can be highly sensitive to the spatial arrangement of surrounding water molecules, leading to pronounced, configuration-dependent bandgap variations. To elucidate the underlying mechanism, we develop predictive ML models trained on a total of 1320 DFT calculated bandgap values to decode the structure-bandgap relationship of NPGs across single and bulk-water environments. To capture structural effects, we compare NPG[37] and hybrid nitrogen-doped NPG (h-NPG),[21] which both have been experimentally realized. A schematic illustration of our ML approach is given in **Figure 1.**

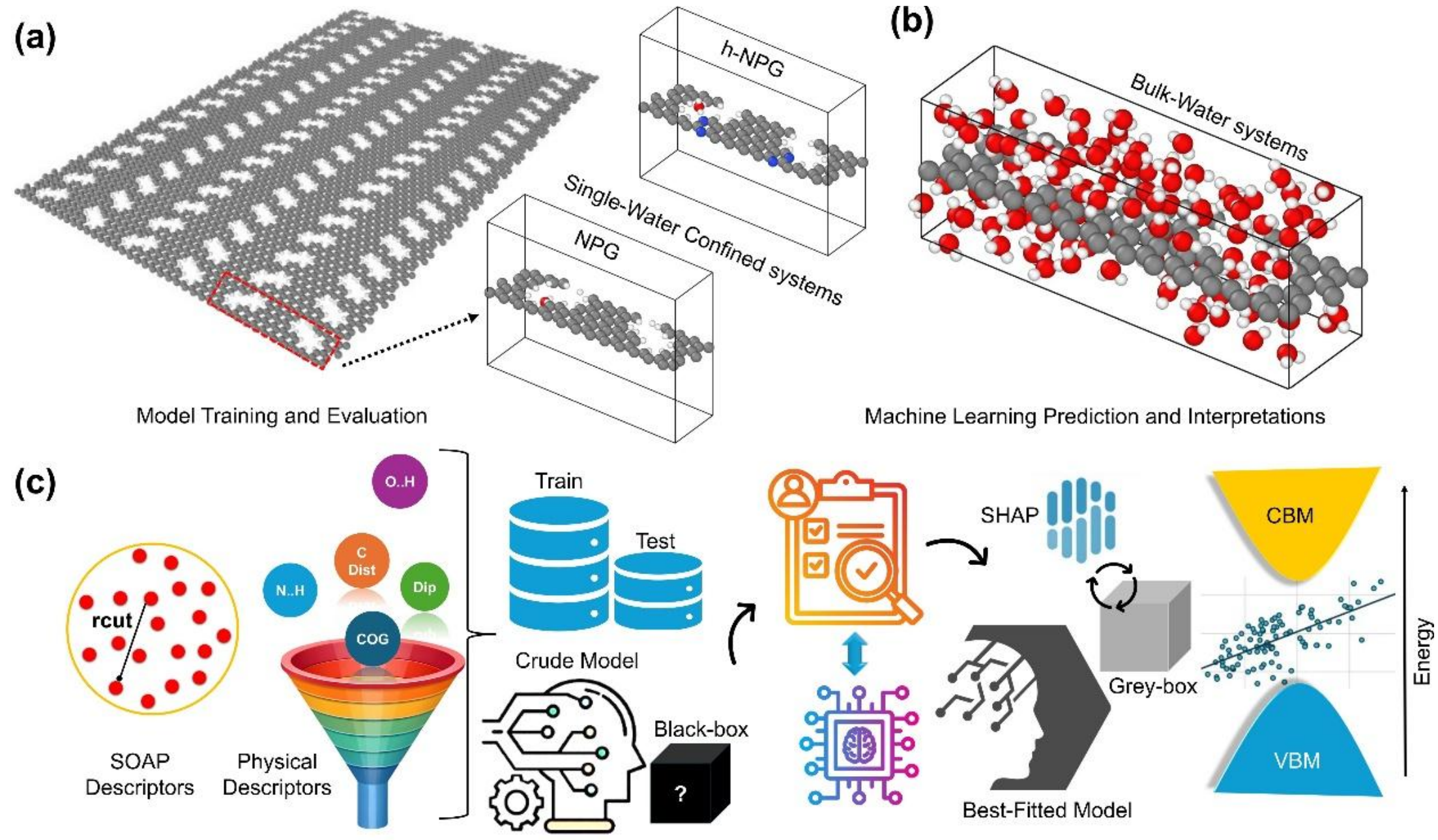

**Figure 1.** (a) Atomic structure of NPG showing the selected unit-cell region used to model single and bulk water NPG/h-NPG systems. Single-water configurations are shown for NPG and h-NPG, (b) a representative bulk-water configuration, and (c) an end-to-end ML framework comprising data generation, model training, prediction, and interpretation. Structural descriptors, including SOAP and physics-informed descriptors, are extracted to encode the relationship between atomic configurations and bandgap energies. Multiple data-driven black-box and physics-informed grey-box models are trained and evaluated, followed by hyperparameter tuning. The best-fitted models are then applied to predict configuration-dependent bandgap energies of previously unseen NPG and h-NPG structures that balance predictive accuracy and interpretability. Model interpretation is performed using Shapley additive explanations (SHAP) to identify key structural and environmental descriptors controlling bandgap modulations.

## RESULTS AND DISCUSSION

### Electronic Structure and Quantum Transport

As a first step, we performed DFT simulations to compare the band structures of periodic unit cells of NPG and h-NPG with a single-water molecule adsorbed at the pore site (**Figure 2a**). The presence of water has a negligible effect on NPG, with its band structure remaining largely unchanged (**Figure 2b**). A direct bandgap of 0.624 eV is observed at the Γ point. The different band dispersions along $\Gamma \rightarrow Z$ and $\Gamma \rightarrow X$ indicate an anisotropic electronic structure, consistent with Ref. 37. In contrast, h-NPG shows a clear response to water. The bandgap decreases from 0.381 eV in the pristine system to 0.241 eV upon water interaction, highlighting its enhanced sensitivity to local hydration. To further understand this behavior, we examine site-dependent interactions by placing water at top, bridge, hollow, and pore sites, as marked in **Figure 2a**. The pore site is energetically most favorable for both NPG and h-NPG, as shown by relative total energies (**Figure S1**). The interaction is stronger in h-NPG, with a binding energy of -0.45 eV, than in NPG, with -0.23 eV, consistent with charge density difference (CDD) results showing enhanced charge redistribution. The stronger interaction in h-NPG can be attributed to enhanced dipole-dopant interactions and hydrogen-bonding interactions with water. Accordingly, the total density of states (TDOS) exhibits more pronounced changes near the Fermi level in h-NPG, reflecting higher electronic sensitivity. Unlike NPG, band structure analysis reveals larger bandgap variations ~ 0.44-0.24 eV depending on adsorption sites in h-NPG, as shown in **Figure S2**. We then investigate density-dependent effects by increasing the number of water molecules per pore from one to three, as can be seen in **Figure 2c**. With increasing water, the bandgap of NPG remains largely unchanged, while h-NPG exhibits a strong, non-monotonic dependence on water density (**Figure S3**), indicating water dipole-dopant electrostatic interactions as a key factor governing bandgap modulations in h-NPG.

The above analyses capture the static electronic response of NPG and h-NPG to water adsorption; however, aqueous environments are inherently dynamic. Thermal fluctuations at room temperature continuously modulate water orientations, hydrogen-bond networks, and interfacial dipoles, which can directly influence the electronic response. To study the dynamics of confined water in NPG and h-NPG for different adsorption sites and water densities, we performed AIMD simulations in the NVT ensemble at 300 K for 1 ps with a 1 fs time step using the Nose-Hoover thermostat, as implemented in SIESTA.[38]

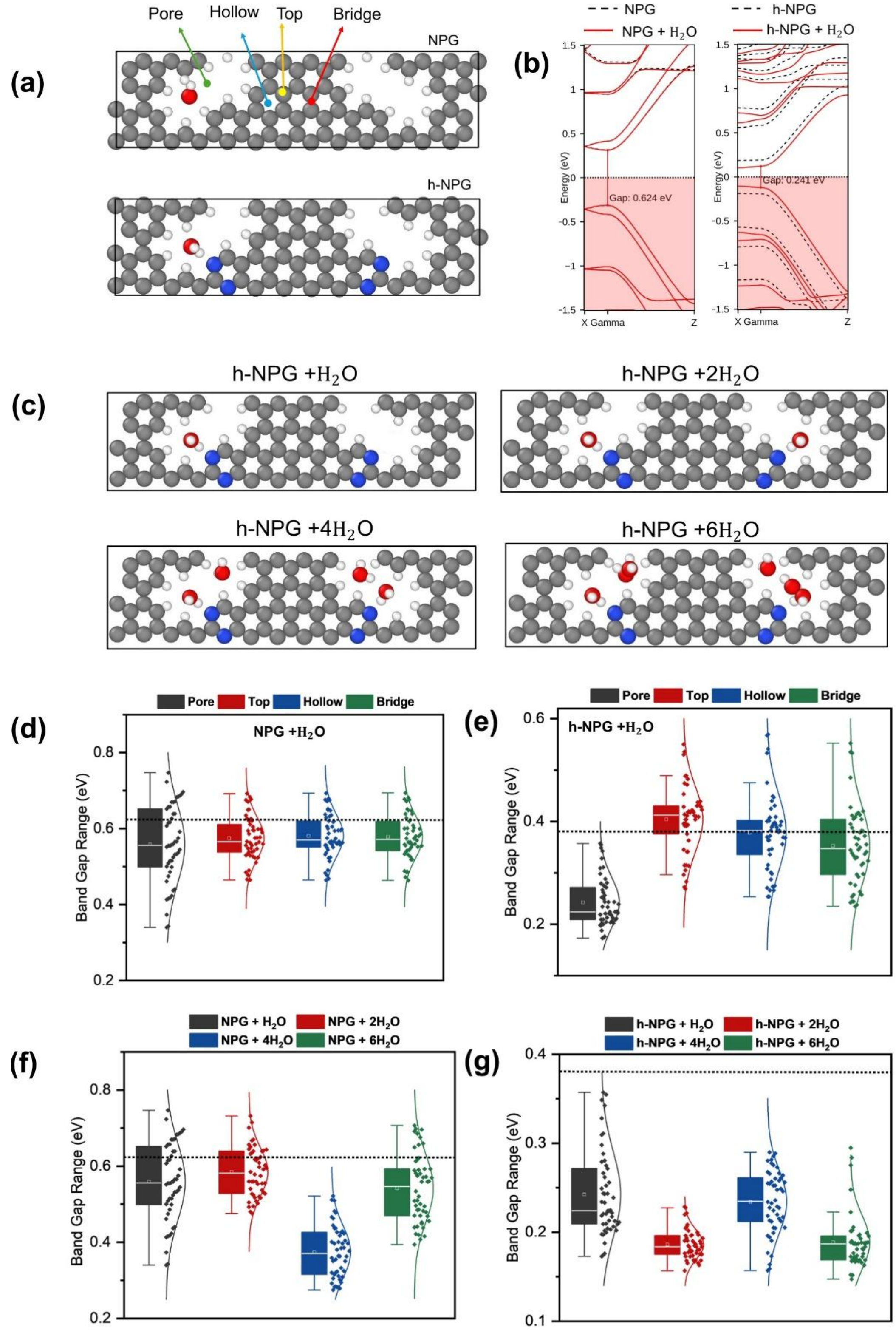

**Figure 2.** (a) Periodic unit cells of NPG and h-NPG with a single-water molecule adsorbed at the pore site, the arrows show different adsorption sites for water, and (b) electronic band structures of NPG and h-NPG in the presence of a single-water molecule. The band dispersions are plotted along the high-symmetry path $X \rightarrow \Gamma \rightarrow Z$ in the Brillouin zone. The black dashed band lines indicate the corresponding band structures of the pristine systems for comparison. (c) Schematic of the optimized configuration of h-NPG with an increasing number of water

molecules (1, 2, and 3 $H_2O$ molecules in each pore). Box normal plots summarizing the statistical distribution of bandgap values for (d) NPG and (e) h-NPG with a single $H_2O$ molecule adsorbed at different sites (pore, top, hollow, bridge). Box-normal plots summarize the statistical distribution of bandgap values for (f) NPG, and (g) h-NPG as a function of increasing number of water molecules. In each case, the boxes represent the interquartile range, the central line denotes the median bandgap, and the whiskers indicate the full range sampled during the trajectory. The black dotted line represents the bandgap of NPG and h-NPG without water. Atom color code: carbon (grey), hydrogen (white), oxygen (red), and nitrogen (blue).

We emphasize that the AIMD simulations are not intended to yield highly converged thermodynamic averages. Instead, their primary purpose is to sample physically realistic and dynamically accessible water configurations representative of ambient conditions. To analyze bandgap fluctuations, 50 equally spaced frames are extracted from each trajectory. For NPG, among the four sites, the pore site is more sensitive to water interaction with a wider distribution in the bandgap (**Figure 2d**). The median bandgap remains virtually the same across all adsorption geometries, indicating weak electrostatic coupling. In contrast, h-NPG exhibits pronounced site-dependent behavior (**Figure 2e**). Water adsorption at the pore site leads to a distinct distribution with the lowest median bandgap. Upon increasing the number of water molecules in NPG, a downward shift in the median bandgap is observed for two water molecules per pore, suggesting cooperative water dipole interactions that enhance the local electrostatic field and thereby reduce the bandgap (**Figure 2f**). In contrast, h-NPG shows consistently narrower bandgap values compared to its pristine counterpart (**Figure 2g**). The smallest spread and lowest bandgap occur for a single water molecule per pore, indicating a stable and well-defined water dipole-dopant interaction in h-NPG.

We also analyze the band structures of NPG and h-NPG in the presence of 114 $H_2O$ molecules, corresponding to a bulk-like water density of ~1.0 g/cm$^3$ within the fixed simulation cell. The initial water configurations are generated using the PACKMOL package.[39] Under bulk-water conditions, the NPG bandgap remains nearly unchanged, whereas in h-NPG, it increases from 0.381 to 0.576 eV (**Figure 3a**).

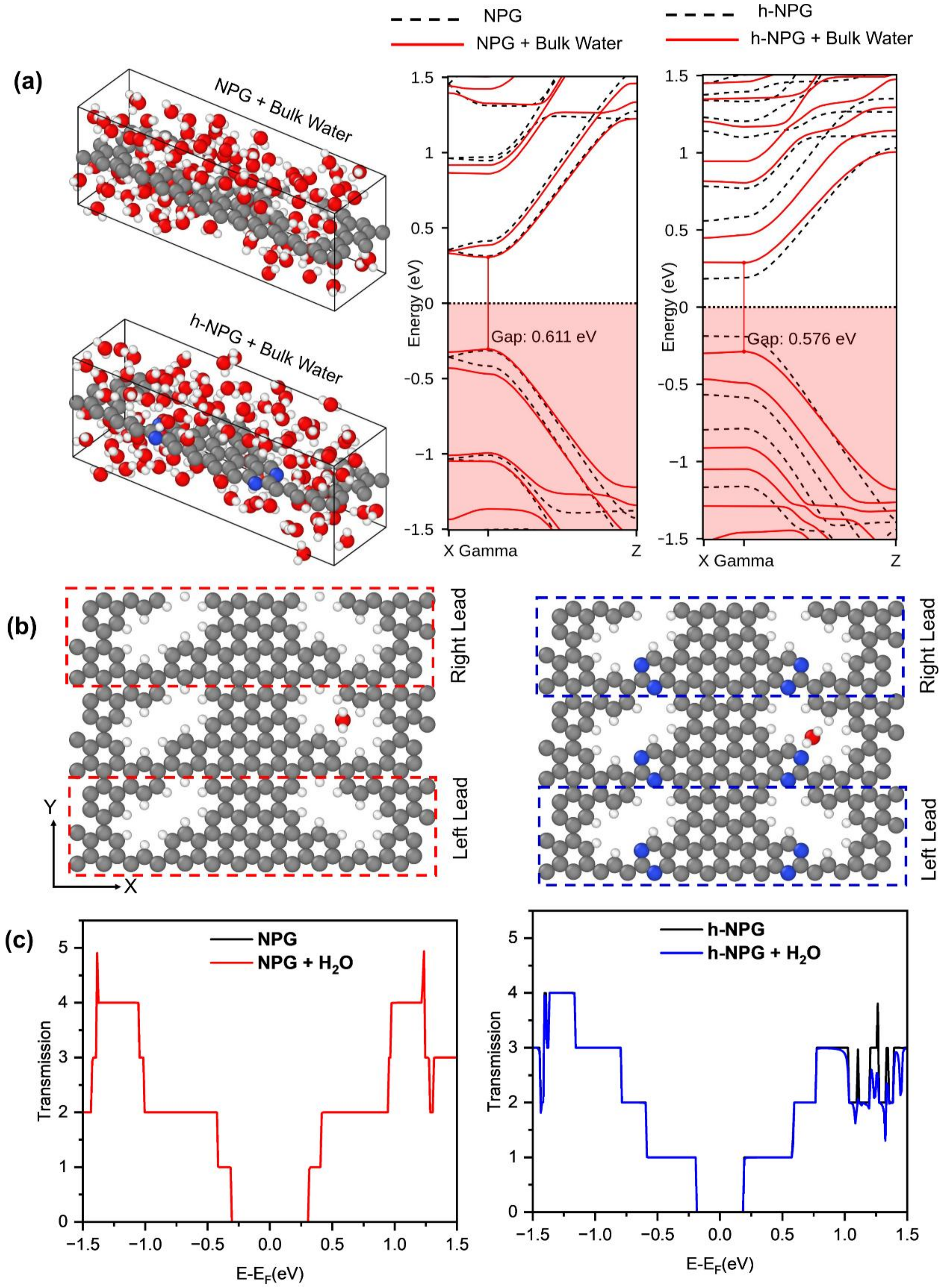

**Figure 3.** (a) Periodic unit cells of NPG and h-NPG under bulk-water at ambient density ($\rho = 1.0\ g\ cm^{-3}$) and corresponding band structures, (b) Two-probe transport geometries of NPG/h-NPG with water adsorbed at the pore site. The dashed boxes indicate the electrode regions used in the transport calculations, and (c) Zero-bias transmission spectra of single-water NPG/h-NPG systems. For comparison, the transmission spectra of corresponding pristine systems are also shown. The Fermi energy is shifted to zero.

This opposite trend compared with the single-water case indicates that the bandgap response of h-NPG is highly sensitive to water density and dipole arrangement. While a single-water molecule reduces the bandgap through a local dipole-dopant interaction, the bulk-water environment introduces collective dipole screening and structural/electrostatic rearrangement around the N-doped pore, leading to bandgap widening. After analyzing water-induced band-structure changes, we investigate quantum transport in the energetically most stable single-water NPG/h-NPG systems (**Figure 3b**) using the nonequilibrium Green's function approach, as implemented in TranSIESTA.[40] The zero-bias transmission spectra of NPG and h-NPG exhibit step-like, quantized features, characteristic of ballistic transport through nanoscale channels (**Figure 3c**). Although water adsorption modulates the bandgap, the transmission near the Fermi level remains nearly unchanged for both NPG and h-NPG. This indicates that the water-induced electronic perturbation mainly shifts or modifies states away from the dominant transport channels, while the conducting states at the Fermi level remain robust.

To better understand local electrostatic perturbations induced by a single-water molecule, we calculate gate-tunable electronic band structures, modeled using the capacitor approach within the SIESTA package.[38] The gate planes are positioned 6 Å above and below the system to simulate top- and bottom-gate effects. The gate carries a charge density of $n = g \times 10^{13}\ e/cm^2$, where $g$ defines the level of gating, such that the system is $n$ doped for $g < 0$ and $p$-doped for $g > 0$. For NPG, gating primarily induces a rigid shift of the Fermi level, with negligible modification of the band structure beyond band filling (**Figure S4**).[41] In contrast, h-NPG exhibits a pronounced gate-dependent response in the single-water configuration, where electrostatic gating opens the bandgap and drives the electronic structure toward that of NPG (**Figure S5**). These observations indicate that localized nitrogen-derived states can be selectively modulated by gating, in contrast to the delocalized carbon states.

These results establish that the electronic response of NPG and h-NPG is strongly influenced by the position, orientation, and density of water molecules, with h-NPG showing a much higher sensitivity due to the presence of nitrogen dopants. However, the structural origin of these bandgap changes is difficult to identify from DFT and AIMD analysis alone. To address this, we develop an interpretable, structure-aware ML framework to predict the bandgap of hydrated NPG/h-NPG systems and identify the key structural descriptors that control the electronic response.

For this purpose, we consider both single-water and bulk-water configurations. From the AIMD trajectories (**Supplementary Videos 1-4**), we extracted 460 frames for the single-water systems and 200 frames for the computationally expensive bulk-water systems, followed by DFT bandgap calculations for each structure. In total, 1320 band structure calculations are performed across NPG/h-NPG systems. The corresponding bandgap histograms are shown in **Figure S6**. The single-water configurations span bandgaps of approximately ~ 0.32-0.75 eV for NPG and ~ 0.17-0.41 eV for h-NPG, whereas the bulk-water configurations span ~ 0.16-0.50 eV for NPG and ~ 0.20-0.68 eV for h-NPG.

For visual understanding, the lowest and highest bandgap structures for each system, together with their relative total energies are shown in **Figure 4**. For single-water configurations, the lowest-bandgap cases appear to be associated with water pointing towards the bridge, particularly close to the N-dopant sites in h-NPG, suggesting stronger local electrostatic coupling (**Figure 4a**). In contrast, the highest-bandgap cases show a different local water arrangement with weaker coupling to the pore/dopant region. For bulk-water NPG, higher-bandgap configuration appears to be associated with stronger local structural or electrostatic perturbations close to the bridge region (**Figure 4b**). Notably, under both single-water and bulk-water environments, the smaller bandgap structures have also smaller relative total energies, indicating that energetically stable water configurations promote stronger water-substrate interactions and more pronounced bandgap modulations.

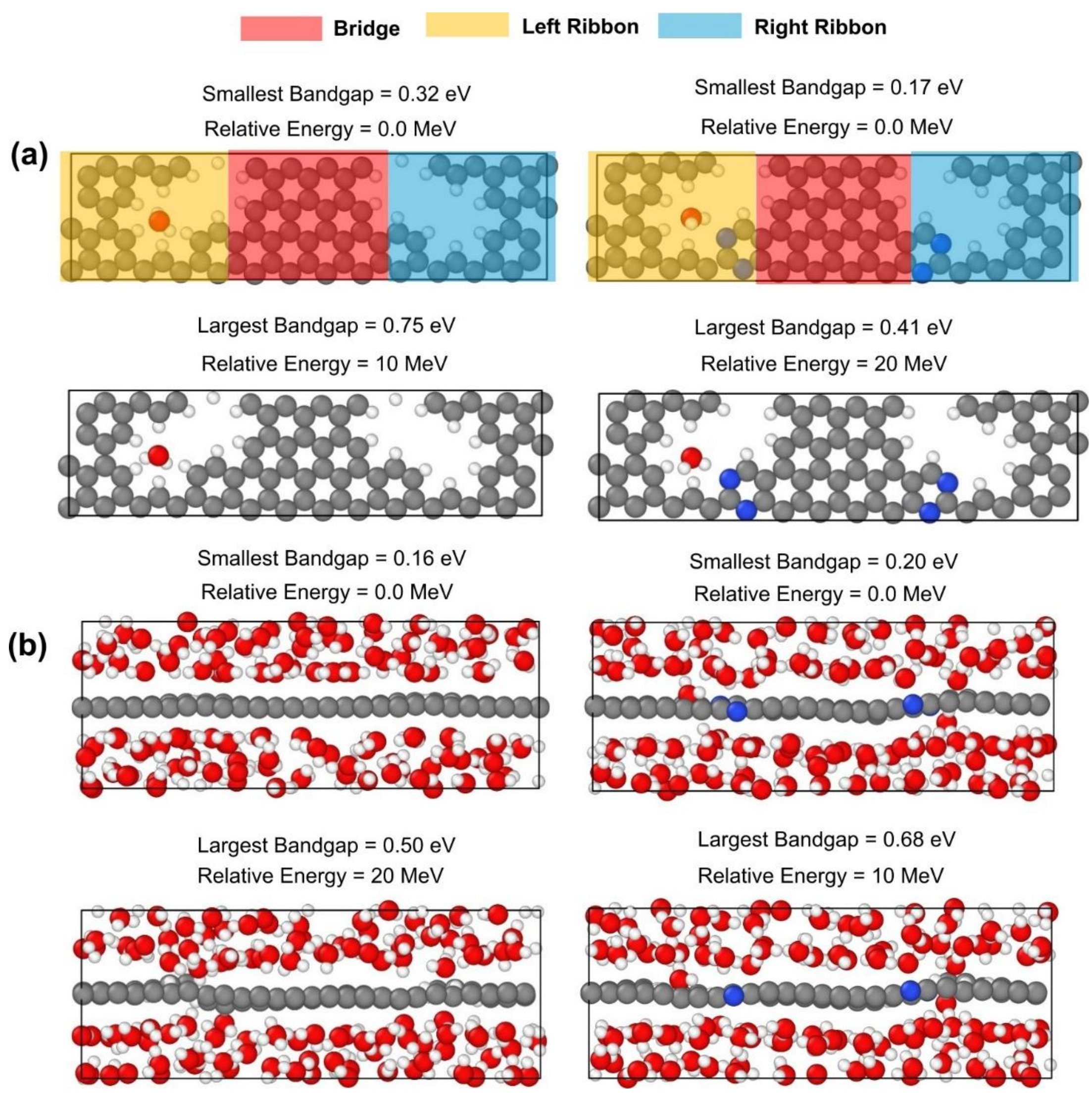

**Figure 4.** The lowest and highest bandgap configurations of NPG and h-NPG in (a) single-water and (b) bulk water environments. Atom color code: carbon (grey), hydrogen (white), oxygen (red), and nitrogen (blue).

## Black-Box ML Predictions

Building on these observations, we next use ML black-box models to test whether the water dependent bandgap variations can be learned from structural descriptors or not. Unlike conventional approaches that rely primarily on structure-based descriptors with different molecular formulas, we use configuration-based descriptors (same molecular formula) that explicitly incorporate local aqueous environment in the predictions.[42–45] Each selected AIMD snapshot is represented using SOAP descriptors ($r_{cut}$ = 6 Å , $n_{max}$= 8, and $l_{max}$= 6), as implemented in the DScribe package.[46] SOAP provides a continuous and differentiable representation of local atomic environments that is invariant to translations and rotations, making it particularly well suited for learning structure-property relationships in atomistic systems. For single-water configurations, this produces 3695 features per structure. For the bulk-water systems, each snapshot contains more than 400 atoms, making the full SOAP representation computationally expensive. Therefore, we use atom-centered SOAP descriptors focused on the NPG/h-NPG framework within a 6 Å cutoff, allowing the local water environment to be captured while keeping the feature size manageable. Because SOAP descriptors are highly expressive, the number of features increases rapidly with the chosen descriptor parameters.

To reduce the computational cost and remove redundant information, we apply principal component analysis (PCA), which is a linear dimensionality reduction technique that transforms the high-dimensional datasets into a smaller set of uncorrelated principal components (PCs) while retaining the most relevant information.[47] Each PC is a linear combination of the original SOAP descriptors, and the number of PCs is chosen based on the point where adding more components does not significantly increase the explained variance. The explained variance measures how much of the dataset's total information (or variability) is captured by each PC. The reduced feature space is then used to generate the final input datasets, comprising 460 data points for each single-water NPG/h-NPG system and 200 data points for each bulk-water NPG/h-NPG system. The datasets are split into 80% training and 20% testing.

To visualize the diversity of local atomic environments and their connection to bandgap, we project the feature space onto a two-dimensional manifold using UMAP (uniform manifold approximation and projection).[48] UMAP finds the nearest neighbors of each point in the high-dimensional space and then constructs a two-dimensional map that preserves these neighborhood relationships as much as possible. In this representation, each point represents an AIMD snapshot, where points that are close together have similar local structures, and the color shows the corresponding bandgap. For single-water NPG, the points form a more dispersed manifold, showing larger variations in local water configurations and a smooth bandgap change along different branches (**Figure 5a**).

In contrast, single-water h-NPG forms a more continuous S-shaped manifold, indicating a more ordered structural evolution with a relatively bigger bandgap distribution. Under bulk-water conditions, NPG exhibits multiple elongated branches corresponding to heterogeneous local environments, with bandgap values varying continuously along each branch, whereas h-NPG forms a small number of well-defined clusters, with narrower bandgap distributions in some regions and broader variation in others, reflecting the combined influence of dopants and collective water interactions (**Figure 5b**).

We then train multiple supervised ML regression models using the PCA-reduced SOAP descriptor space to predict bandgaps for previously unseen NPG and h-NPG configurations under both single- and bulk-water environments. The selected algorithms include Gaussian process regression (GPR), Kernel Ridge Regression (KRR), Random Forest Regression (RFR), and Extreme Gradient Boosting Regression (XGBR). Each model is trained using optimized hyperparameters, as summarized in **Table S1**. Among these models, GPR provided strong predictive performance in all four datasets, as summarized in **Table S2**, and is therefore selected for further predictions. To further improve the GPR model performance and remove less informative features, we apply greedy recursive feature elimination to the PCA-reduced SOAP feature set.[49] In this procedure, the descriptors are iteratively evaluated, and the least useful features are removed until an optimized subset of descriptors is obtained. Using this optimized descriptor subset, we trained GPR models to predict the DFT-calculated bandgaps of previously unseen NPG and h-NPG configurations.

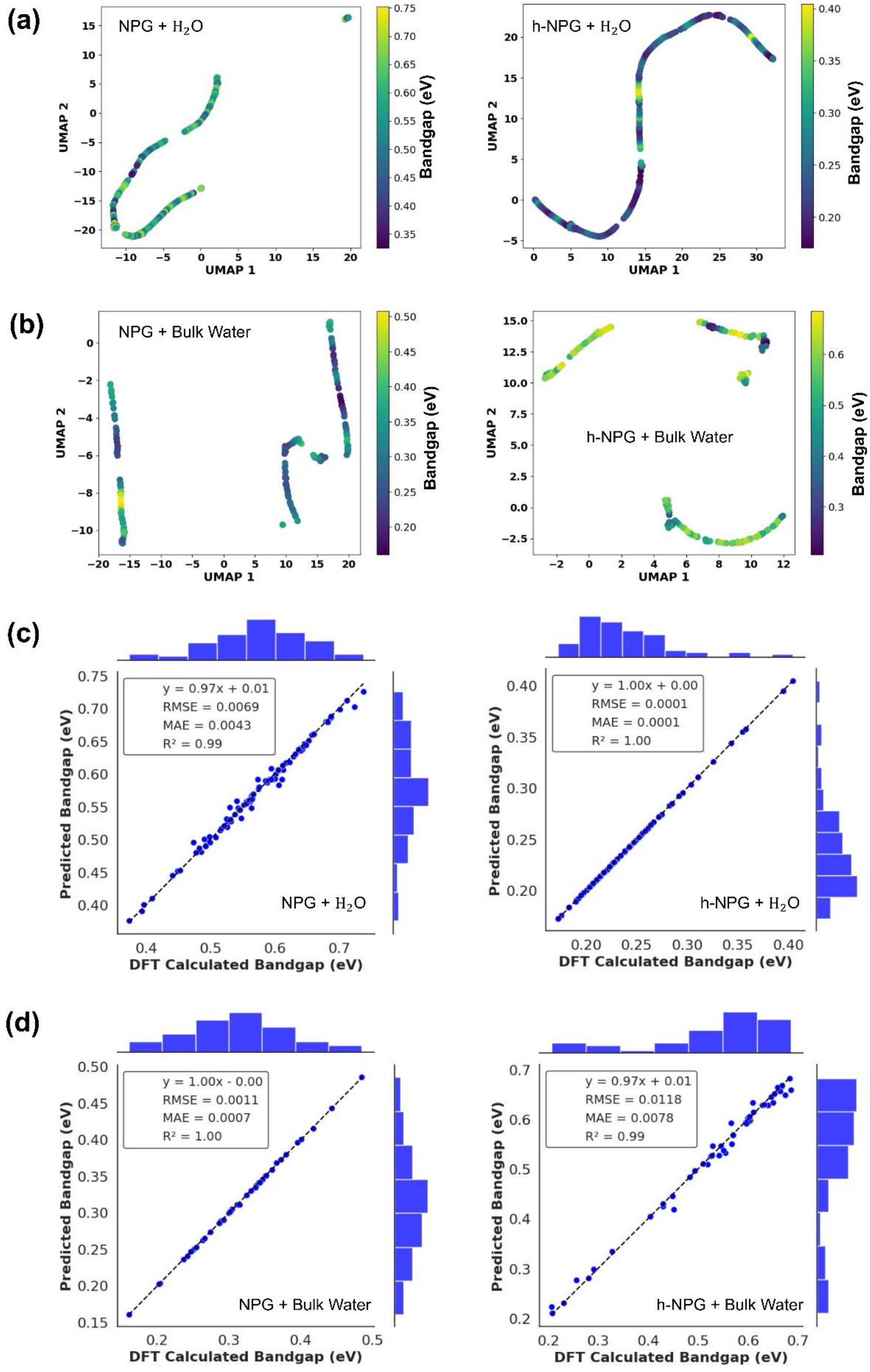

**Figure 5.** ML prediction of bandgaps from SOAP-based structural descriptors. Two-dimensional UMAP embeddings of the reduced feature space for (a) single-water NPG/h-NPG and (b) bulk-water NPG/h-NPG configurations, colored by DFT-calculated bandgap energies. The axes UMAP1 and UMAP2 are latent embedding coordinates generated by the algorithm. Marginal histogram parity plots comparing GPR-predicted and DFT-calculated bandgaps for (c) single-water and (d) bulk-water configurations, with linear fits and error metrics showing the accuracy of the best-performing model.

The improvement after feature selection is evident from the parity plots and marginal histograms, where the predicted bandgaps closely follow the DFT-calculated values and most data points lie near the ideal diagonal line (**Figure 5c-d**). Notably, in the bulk-water configurations also, the model achieved near-DFT performance ($R^2$ ~ 0.99-1.00, RMSE ~ 0.012-0.001 eV). For the final GPR model, predictions are also reported with 95% confidence intervals obtained from the predictive standard deviation of the Gaussian process (**Figure S7**). The uncertainty bars provide an additional measure of model reliability, indicating where the model is more or less confident in its bandgap predictions. To avoid relying on a single train–test split, we further evaluate the GPR model using repeated 10-fold cross-validation. The mean RMSE values remain low across all systems, with cross validation RMSE values of 0.0164 ± 0.0079 eV for single-water NPG, 0.0009 ± 0.0012 eV for single-water h-NPG, 0.0050 ± 0.0034 eV for bulk-water NPG, and 0.0320 ± 0.0138 eV for bulk-water h-NPG. These small mean errors and standard deviations indicate that the models are stable across different train-test splits. The learning curves further support the robustness and generalizability of the model, showing that the cross-validation RMSE decreases as the training-set size increases and gradually approaches a stable value (**Figure S8**).

**Grey-Box ML Predictions**

Motivated by the black-box ML results, we next develop a grey-box ML framework to identify the key structural and electronic descriptors controlling bandgap changes. Before analyzing the more complex bulk-water systems, we first select physically meaningful descriptors for the simpler single-water NPG and h-NPG systems. These descriptors provide a baseline for designing descriptors for the bulk-water systems. A detailed description of the extracted features and their mathematical formulations is provided in **Supporting Note 1**, while concise summary is presented in **Table 1**.

**Table 1.** Description of selected features for bandgap prediction of single-water NPG/h-NPG systems.

| S. No. | Feature label | Description |
|---|---|---|
| **1** | $COG_X(H_2O)$ | *x*-component of the water center of geometry |
| **2** | $COG_y(H_2O)$ | *y*-component of the water center of geometry |
| **3** | $COG_z(H_2O)$ | *z*-component of the water center of geometry |
| **4** | $d_{min}(O_w, H_{pore})$ | Minimum distance between the water oxygen and pore hydrogen atoms |
| **5** | $d_{min}(H_w, N)$ | Minimum distance between water hydrogen atoms and N-dopant atoms |
| **6** | $\mu_X(H_2O)$ | *x*-component of the water dipole moment |
| **7** | $\mu_y(H_2O)$ | *y*-component of the water dipole moment |
| **8** | $\mu_z(H_2O)$ | *z*-component of the water dipole moment |
| **9** | $\vert\mu(H_2O)\vert$ | Magnitude of the water dipole moment |
| **10** | $\theta_{\mu,z}$ | Angle between the water dipole moment vector and the surface normal direction |
| **11** | $d_{min}(H_2O, C)$ | Minimum distance between the water center of geometry and framework carbon atoms |
| **12** | $d_{max}(H_2O, C)$ | Maximum distance between the water center of geometry and framework carbon atoms |
| **13** | $\sigma_d\ [d(H_2O, C)]$ | Standard deviation of distances between the water center of geometry and framework carbon atoms |

Using these physics-informed descriptors, we prepared input datasets containing 460 data points for the single-water NPG and h-NPG systems. Similar to the black-box ML analysis, greedy recursive feature elimination is employed to identify the optimal set of physics-informed descriptors while removing redundant information. The grey box GPR results for the single-water NPG and h-NPG systems are shown in **Figure 6**. The model accurately reproduces the DFT-calculated bandgaps for both systems. For NPG, the GPR predicted values follow the DFT trend with $R^2$ = 0.94, although a few deviations are observed at higher bandgap values (**Figure 6a**). For h-NPG, the prediction is nearly ideal, with $R^2$ = 1.00 and a test RMSE of 0.0007 eV. The reliability of the grey-box GPR models is further evaluated using repeated 10-fold cross-validation. The mean cross-validation RMSE values are 0.0254 ± 0.0087 eV for NPG and 0.0020 ± 0.0014 eV for h-NPG. These cross-validation results indicate that the predicted bandgaps remain close to the DFT-calculated values across different train-test splits. To interpret the grey-box GPR models, we calculated Shapley additive explanations (SHAP) values for the selected physics-informed descriptors. SHAP analysis assigns an importance value to each descriptor based on its contribution to the model prediction. The SHAP analysis reveals the key physical descriptors controlling the bandgap response (**Figure 6b**). For single-water NPG, the two most important descriptors are $\Theta_{\mu,z}$ and $\sigma_d$ $[d(H_2O, C\ )]$, which together account for 38.5% of the total feature contribution. This indicates that the bandgap of NPG is primarily governed by the orientation of the water dipole and the distribution of water–carbon distances. For h-NPG, the dominant descriptors are $COG_X(H_2O)$ and $\mu_z(H_2O)$, which together account for 40.2% of the total feature contribution. This suggests that the bandgap response of h-NPG is sensitive to both the lateral position of the water molecule and its out-of-plane dipole component. Building on these results, we next extend the grey-box descriptor space to the bulk water NPG/h-NPG systems. Compared with the single-water case, the bulk-water environment requires additional descriptors to capture collective water effects, water–substrate interactions, and water–water interactions. To define the physically relevant water-substrate interactions, a cutoff distance of 5.0 Å is used. In total, 38 structural descriptors are generated for the bulk-water systems. The detailed physical motivation and mathematical formulation of each descriptor are provided in **Supporting Note 2**, while selected feature abbreviations and short descriptions are presented in **Table 2**. For both datasets, each containing 200 data points, we apply greedy recursive feature elimination, and the optimized feature subsets are then used to train the GPR models for the bandgap prediction of corresponding test datasets.

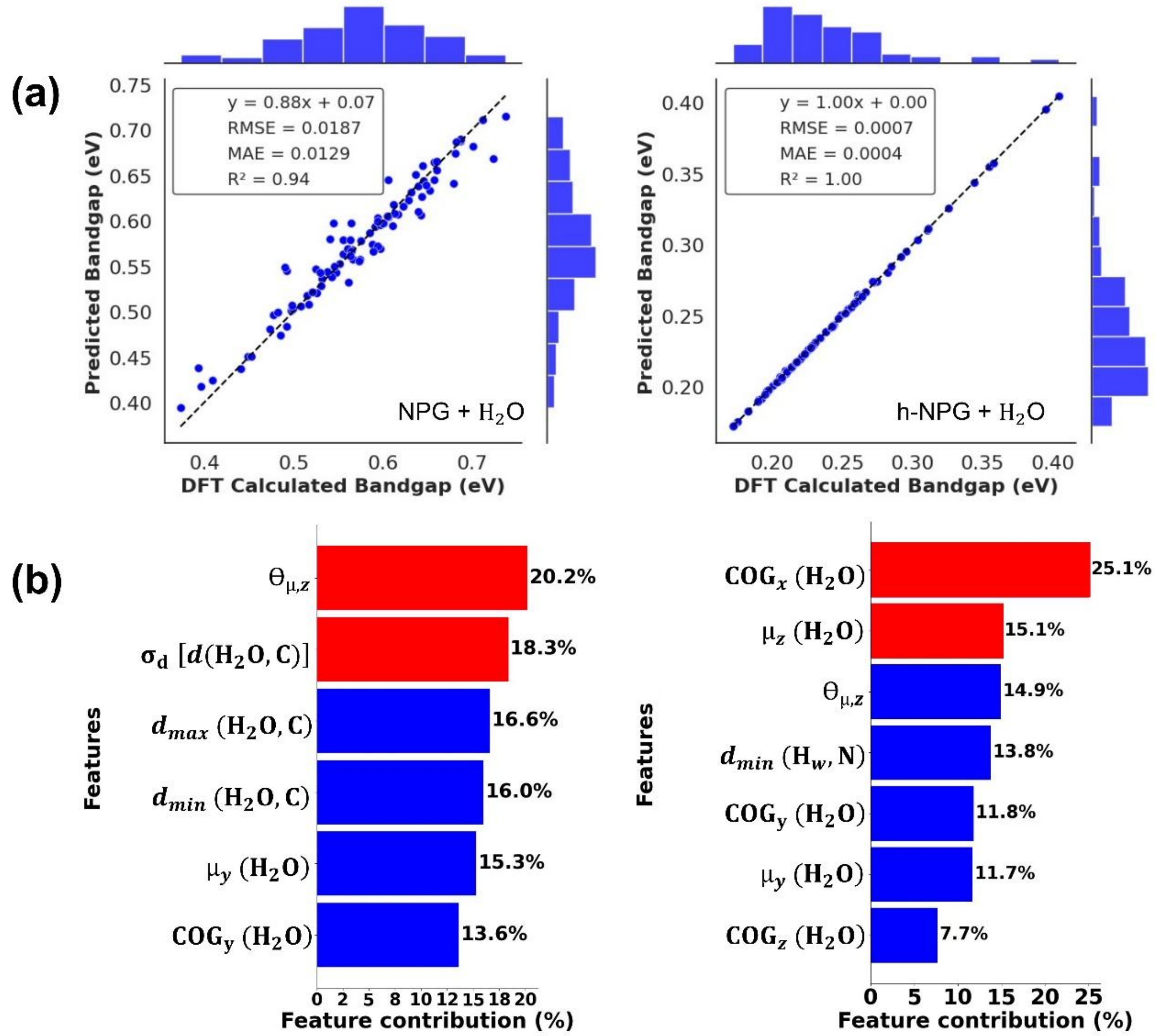

**Fig. 6** Grey-box ML prediction and feature-importance analysis for single-water NPG and h-NPG systems. (a) Marginal histogram parity plots comparing GPR-predicted bandgaps with DFT-calculated bandgaps for NPG $+ H_2O$ and h-NPG $+ H_2O$ using physics-informed descriptors. (b) SHAP-based global feature-importance plots showing the relative contribution of the selected descriptors to the bandgap prediction for each system.

The grey-box GPR results for the bulk-water NPG and h-NPG systems are shown in **Figure 7a**. For bulk-water configurations, the GPR models achieve high predictive accuracy, with $R^2 = 1.00$ and RMSE $= 0.0021$ eV for NPG, and $R^2 = 0.99$ and RMSE $= 0.0123$ eV for h-NPG. These results show that the selected physics-informed descriptors can capture the underlying physical mechanism behind bandgap variations in the bulk-water NPG/h-NPG systems. The repeated 10-fold cross-validation RMSE values are $0.0076 \pm 0.0052$ eV for NPG and 0.0261 $\pm$ 0.0089 eV for h-NPG. These validation results across different train-test splits are also in good agreement with the DFT-calculated values indicating that the grey-box GPR models are not overfitted. The SHAP feature-importance results identify $\Delta\ |\mu|_{bridge-left/right}$ as the most important descriptor in bulk-water NPG, contributing 22.1% of the total feature contribution (**Figure 7b**).

**Table 2** Description of selected features for bandgap prediction of bulk-water NPG/h-NPG systems.

| S. No. | Feature label | Description |
|---|---|---|
| **1** | $mean[COG_{x/y/z}\ (H_2O)]$ | Mean *x*, *y*, and *z* coordinates of water centers of geometry |
| **2** | $\sigma_d[COG_{x/y/z}\ (H_2O)]$ | Standard deviation of the water center-of-geometry positions along the *x*, *y*, and *z* directions |
| **3** | $\mu^{tot}_{x/y/z}$ | *x*, *y*, and *z* components of the total dipole moment of the water layer |
| **4** | $\lvert\mu^{tot}\rvert$ | Magnitude of the total dipole moment of the water layer |
| **5** | $mean[\lvert\mu\rvert]$ | Mean magnitude of individual water dipole moments |
| **6** | $\sigma_d[\lvert\mu\rvert]$ | Standard deviation of individual water dipole magnitudes |
| **7** | $mean[\theta_{\mu,z}]$ | Mean angle between individual water dipole moments and the surface normal direction |
| **8** | $\sigma_d[\theta_{\mu,z}]$ | Standard deviation of water dipole orientations relative to the surface normal direction |
| **9** | $min\ [d(O_w, H_w)]$ | Minimum intermolecular O···H distance between water molecules |
| **10** | $mean\ [d(O_w, H_w)]$ | Mean intermolecular O···H distance between water molecules |
| **11** | $min\ [d(N, H_w)]$ | Minimum distance between N-dopant atoms and water hydrogen atoms |
| **12** | $mean\ [d(N, H_w)]$ | Mean distance between N-dopant atoms and water hydrogen atoms |
| **13** | $N_{OH}$ | Number of intermolecular O···H contacts within the cutoff radius $r_c$ |
| **14** | $N_{NH}$ | Number of N···H contacts between N-dopant atoms and water hydrogens within the cutoff radius $r_c$ |
| **15** | $\mu^{bridge}_{x/y/z}$ | x, y, and z components of the total dipole moment of water molecules near the bridge region |
| **16** | $\lvert\mu^{bridge}\rvert$ | Magnitude of the total dipole moment of water molecules near the bridge region |
| **17** | $mean\ [\lvert\mu^{bridge}\rvert]$ | Mean magnitude of individual water dipole moments near the bridge region |
| **18** | $\sigma_d\ [\lvert\mu^{bridge}\rvert]$ | Standard deviation of individual water dipole magnitudes near the bridge region |
| **19** | $\mu^{left/right}_{x/y/z}$ | *x*, *y*, and *z* components of the total dipole moment of water molecules in the combined left/right ribbon regions |
| **20** | $\lvert\mu^{left/right}\rvert$ | Magnitude of the total dipole moment of water molecules in the combined left/right ribbon regions |

| 21 | $mean\ [\lvert\mu^{left/right}\rvert]$ | Mean magnitude of individual water dipole moments in the combined left/right ribbon regions |
|---|---|---|
| 22 | $\sigma_d\ [\lvert\mu^{left/right}\rvert]$ | Standard deviation of individual water dipole magnitudes in the combined left/right ribbon regions |
| 23 | $\Delta\ \mu_{x/y/z}$ | Difference in total dipole components between the bridge region and in the combined left/right ribbon regions |
| 24 | $\Delta\ \lvert\mu\rvert^{bridge-left/right}$ | Difference in total dipole magnitude between the bridge region and in the combined left/right ribbon regions |
| 25 | $\Delta mean\ [\lvert\mu\rvert]^{bridge-left/right}$ | Difference in mean individual water dipole magnitude between the bridge region and in the combined left/right ribbon regions |
| 26 | $\Delta\sigma_d\ [\lvert\mu\rvert]^{bridge-left/right}$ | Difference in dipole-magnitude variability between the bridge region and in the combined left/right ribbon regions |

This indicates that in NPG, the bandgap is mainly governed by a ribbon-resolved imbalance in the collective water-dipole response. Such an imbalance creates local variations in the electrostatic potential between the bridge and left/right ribbon regions, thereby modulating the bandgap. In addition, the spatial geometry of water molecules along the confinement direction and the ribbon-resolved water dipoles also play an important role in modulating the bandgap. For the bulk-water h-NPG system, the dominant descriptors are $\lvert\mu^{left/right}\rvert$, $\sigma_d[COG_x(H_2O)]$ and $mean\ [COG_x(H_2O)]$, which together account for 40.2% of the total feature contribution. These descriptors indicate that the bandgap of h-NPG is strongly influenced by the dipole response of water in the combined left/right ribbon regions and by the lateral distribution of water molecules along the h-NPG surface.

To evaluate the reliability and transferability of our developed grey-box ML framework, we analyzed the predictive confidence intervals and learning curves of physics-informed GPR models across single- and bulk-water NPG/h-NPG systems. The 95% confidence intervals remained narrow for most test predictions, indicating a stable predictive performance (**Figure S9a-b**). Slightly larger uncertainty is observed for the single-water NPG system, consistent with its broader configurational variability. The bulk-water models showed more stable predictions with relatively small uncertainty, suggesting that the collective water descriptors provide a robust representation of the hydration environment (**Figure S9c-d**). The learning curves further confirmed this behavior, showing a systematic decrease in prediction error with increasing training-set size and convergence toward a stable error plateau (**Figure S10**). These results indicate that the grey-box models are not overfitted and that the selected physics-informed descriptors capture transferable structure–property relationships across single-water and bulk-water NPG/h-NPG systems. Thus, the proposed framework provides both predictive accuracy and physical interpretability, offering a generalizable route for understanding and rapidly screening water-controlled electronic properties in nanoporous graphenes and related two-dimensional materials.

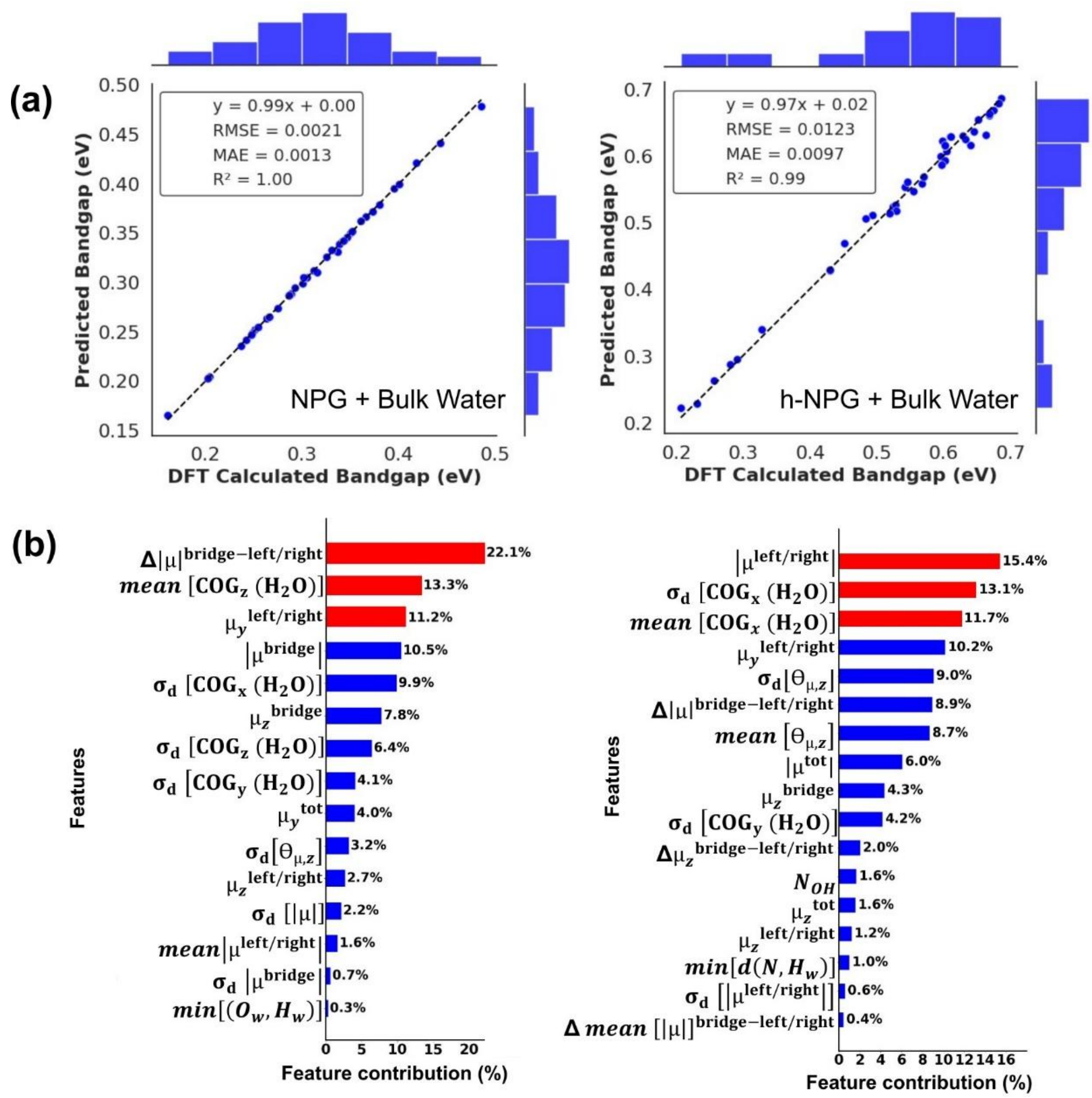

**Figure 7.** (a) Marginal histogram parity plots comparing GPR-predicted bandgaps with DFT-calculated bandgaps for NPG + Bulk water and h-NPG + Bulk water using physics-informed descriptors and (b) SHAP-based global feature-importance plots showing the relative

contribution of the selected physics-informed descriptors to the bandgap prediction for NPG/h-NPG system.

## CONCLUSION

We present an interpretable ML framework that integrates DFT, AIMD, and physics-informed descriptors to rapidly predict the bandgaps of NPG and h-NPG in aqueous environments with near-DFT accuracy. Our results show that the electronic structure of nanoporous graphenes can be sensitive to the position, orientation, and density of water molecules. This effect is particularly pronounced in h-NPG, where nitrogen dopants introduce enhanced sensitivity to local water configurations through water dipole-dopant interactions. The predicted bandgaps vary substantially with the water environment, ranging from ~ 0.32-0.75 eV for NPG and ~ 0.17-0.41 eV for h-NPG in single-water configurations, and from ~ 0.16-0.50 eV for NPG and ~ 0.20-0.68 eV for h-NPG in bulk water. We find that water is not merely a passive surrounding medium but an active tuning parameter for the electronic properties of nanoporous graphenes. To uncover the underlying mechanism, we develop SOAP-based black-box and physics-informed grey-box ML models. Black-box models achieve near-DFT accuracy for bandgap predictions across both single and bulk-water environments. Beyond predictive accuracy, this study underscores the role of explainable ML as a bridge between atomistic simulations and physical understanding of nanomaterials under aqueous environments. By identifying physically meaningful descriptors, our framework establishes a direct link between molecular-scale hydration structure and macroscopic electronic response. Analyses identify water dipole orientation, water-substrate distance, water center-of-geometry, and ribbon-resolved dipole moments as the dominant factors controlling bandgap modulation across NPG and h-NPG systems. We believe that the proposed ML framework can be reliably extended to other 2D materials for rapid bandgap prediction in aqueous environments. Although the physically informed feature space may need adjustment for different materials, it offers a solid starting point for exploring and predicting bandgaps beyond NPGs with a better understanding of underlying physics, facilitating the targeted design of 2D materials with tailored electronic properties.

## METHODS

All first-principles calculations, including ab initio molecular dynamics (AIMD) simulations, are performed within the framework of density functional theory (DFT) using SIESTA version 5.2.0.[38] Core–valence interactions are described using Troullier–Martins norm-conserving pseudopotentials,[50] generated for C, N, O, and H atoms. Valence electrons are expanded in a numerical atomic orbital basis using a double-$\zeta$ polarized (DZP) basis set for all atomic species. An energy shift of 0.15 eV is applied to define the confinement radius of the basis orbitals. The exchange–correlation energy is treated using a consistent-exchange van der Waals density functional (vdW-DF),[51] which provided an accurate description of dispersion interactions critical for graphene–water interfaces. The real-space integration grid is defined by a mesh cutoff of 400 Ry. Convergence tests confirmed that this value yielded total energy variations below 0.1 meV/atom. Structural relaxations are carried out using a conjugate-gradient algorithm until the maximum residual force on each atom is below 0.01 eV/Å. Total energy convergence is set to $10^{-4}$ eV. Periodic boundary conditions are applied along the $x$ and $y$ directions, while a vacuum spacing of 24 Å is introduced along the $z$ direction to avoid spurious interactions between periodic replicas. Brillouin zone sampling is performed using a Monkhorst–Pack $k$-point mesh of 1 × 4 × 1. Convergence tests with denser meshes showed negligible changes in total energies and no variation in bandgap values.

The interaction energy ($E_i$) between the NPG/h-NPG system and the single-water molecule is computed using the following expression:

$$E_i = [E_{NPG/h-NPG+\mathrm{H_2O}} - (\mathrm{E}_{NPG/h-NPG} + \mathrm{E}_{\mathrm{H_2O}})] \quad (1)$$

where $E_{NPG/h-NPG+\mathrm{H_2O}}$ is the total optimized energy of the NPG/h-NPG system with a water molecule placed inside the pore, and $\mathrm{E}_{NPG/h-NPG}$ and $\mathrm{E}_{\mathrm{H_2O}}$ are the total optimized energies of the isolated NPG/h-NPG system and the isolated water molecule, respectively.

The charge density difference (CDD) is evaluated to analyze charge redistribution upon interaction and is defined as:

$$\Delta\rho(r) = [\rho_{NPG/h-NPG+\mathrm{H_2O}}(r) - (\rho_{NPG/h-NPG}(r) + \rho_{\mathrm{H_2O}}(r))] \quad (2)$$

where $\rho_{NPG/h-NPG+\mathrm{H_2O}}(r)$ is the total charge density of the combined NPG/h-NPG–water system, and $\rho_{NPG/h-NPG}(r)$ and $\rho_{\mathrm{H_2O}}(r)$ are the charge densities of the isolated NPG/h-NPG system and isolated water molecule, respectively, all evaluated at the optimized geometry of the combined system.

To calculate the transport properties, DFT combined with the non-equilibrium Green's function (DFT–NEGF) approach is employed, as implemented in TranSIESTA version 5.2.0.[40] Machine-learning analyses are performed in Python using the open-source scikit-learn library. SOAP descriptors are generated with DScribe[46] and used as input features in the black-box ML models. Hyperparameters are optimized using RandomizedSearchCV,[52] and greedy recursive feature elimination is applied for feature selection within scikit-learn.[53] All workflows are executed in Google Colab (https://colab.research.google.com/).

## ASSOCIATED CONTENT

***Supporting Information:**

Electronic Properties of NPG/h-NPG + $H_2O$; Band Structures with Single Water; Band Structures with an Increasing Number of Water Molecules; Electrostatic gating of Single-Water NPG System; Electrostatic gating of Single-Water h-NPG System; Bandgap Histograms under Single and Bulk-water Environment; Tuned Hyperparameters and Model Performance; Confidence Intervals for Black-box GPR Models; Learning Curves for Black-box GPR

Models; Supporting Note 1; Supporting Note 2; Confidence Intervals for Grey-box GPR Models; Learning Curves for Grey-box GPR Models

***SUPPORTING VIDEOS**

- Supplementary Video 1: AIMD trajectory of NPG under a single-water environment.
- Supplementary Video 2: AIMD trajectory of NPG under a bulk-water environment.
- Supplementary Video 3: AIMD trajectory of HNPG under a single-water environment.
- Supplementary Video 4: AIMD trajectory of HNPG under a bulk-water environment.

**ACKNOWLEDGEMENTS**

S.M. acknowledges financial support from Novo Nordisk Foundation (BIONWIRE, grant NNF23OC0084494). A.E.A.M. is funded by Independent Research Fund Denmark, Grant 0.46540/3103-00229B. V.R. acknowledges Novo Nordisk Foundation (MOTIFS, grant NNF23OC0086816). We acknowldege DTU Computing Center for providing computational resources, http://dx.doi.org/10.48714/DTU.HPC.0001.

**COMPETING INTERESTS**

The authors declare no competing interests.

**CODE AND DATA AVAILABILITY**

The datasets and Python scripts used to train, test, and evaluate the machine-learning models developed in this work for predicting the bandgap energies of NPG/h-NPG systems under single- and bulk-water environments have been deposited in a public GitHub repository: https://github.com/Nanoelectronics-Theory-DTU/AquaNPG-ML-Bandgap-Prediction.git

**AUTHOR CONTRIBUTIONS**

M.B. Designed and supervised the project. S.M. performed the computational calculations, developed ML models, and wrote the original draft. A.E.A.M. and V.R. assisted in data analysis. A.E.A.M. and V.R. contributed equally to this work. All authors contributed to the discussions and editing of the manuscript.

**Table of Contents**

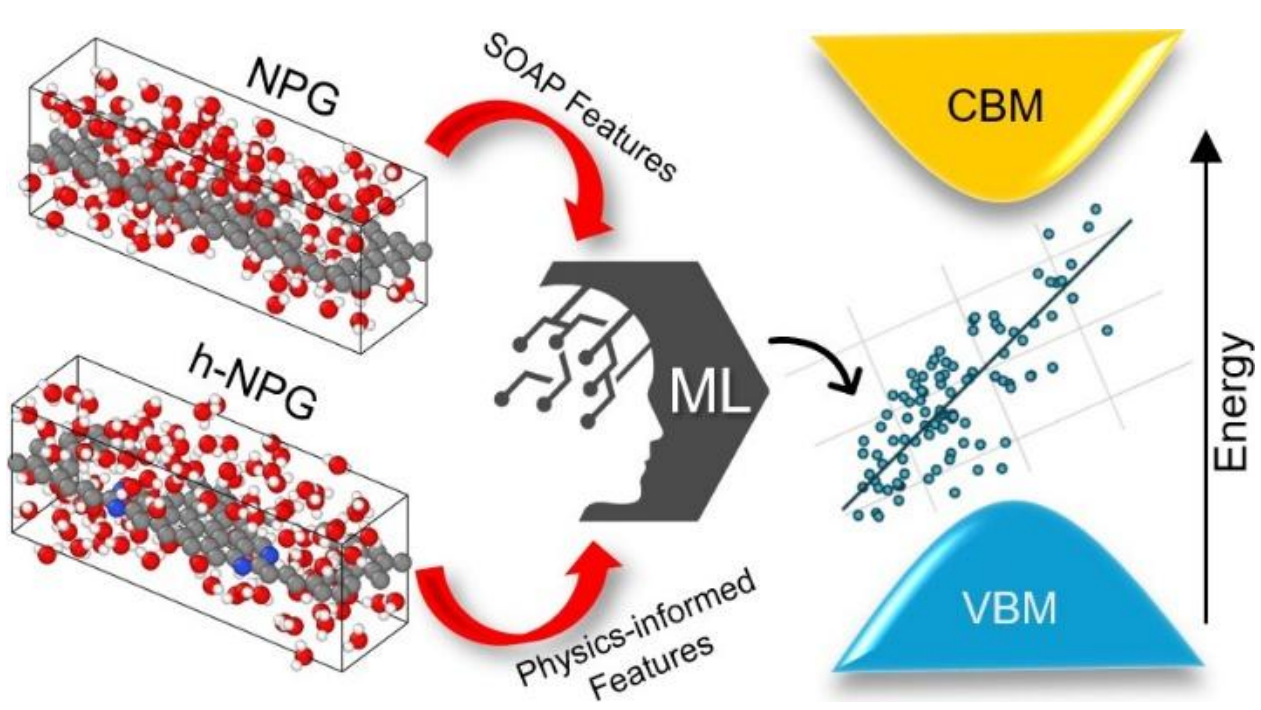

## Supporting Information

# Machine Learning Bandgap Prediction of Nanoporous Graphenes with Water

Sneha Mittal, *[, †] Alan E. Anaya Morales, [†, ‡] Victor Rosendal, [†, ‡] Mads Brandbyge *[, †]
[†]Department of Physics, Technical University of Denmark, 2800 Kgs., Lyngby, Denmark
[‡]These authors contributed equally to this work.
*Corresponding Author. E-mail: snemi@dtu.dk (S.M.); mabr@dtu.dk (M.B.)

**Table of Contents** **Pages**

## 1. Electronic Properties of NPG/h-NPG + $H_2O$

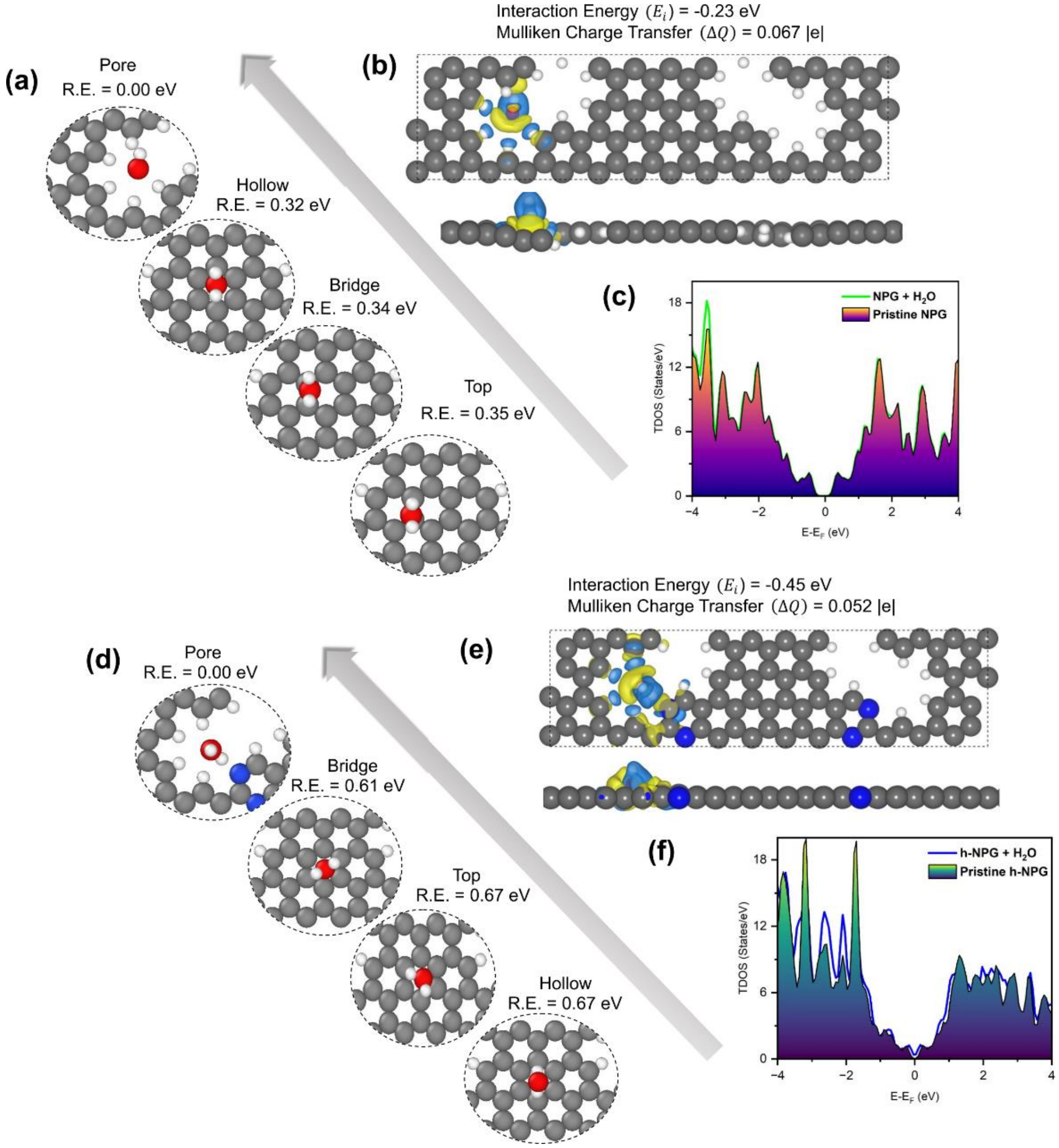

**Figure S1**: (a) Optimized geometries of NPG with a water molecule adsorbed at different sites and the corresponding relative total energy (R.E.) values, (b) charge density difference (CDD) plot for NPG + $H_2O$ system in its energetically most favorable pore configuration, along with the corresponding interaction energy ($E_i$) and Mulliken charge transfer values, (c) total density of states (TDOS) of NPG with and without water molecule, (d) optimized geometries of h-NPG with a water molecule adsorbed at different sites and the corresponding R.E. values, (e) CDD plot of the h-NPG + $H_2O$ system in its energetically most favorable pore configuration, along with the corresponding $E_i$ and Mulliken charge-transfer values, and (f) TDOS of h-NPG with and without an adsorbed water. The Fermi energy level is shifted to zero. The isosurface value is 0.005 $e\text{Å}^{-3}$. Yellow and blue colors in the CDD plot represent charge depletion and charge accumulation, respectively. Atom color code: carbon (grey), hydrogen (white), and oxygen (red).

## 2. Band Structures of NPG/h-NPG with Single Water

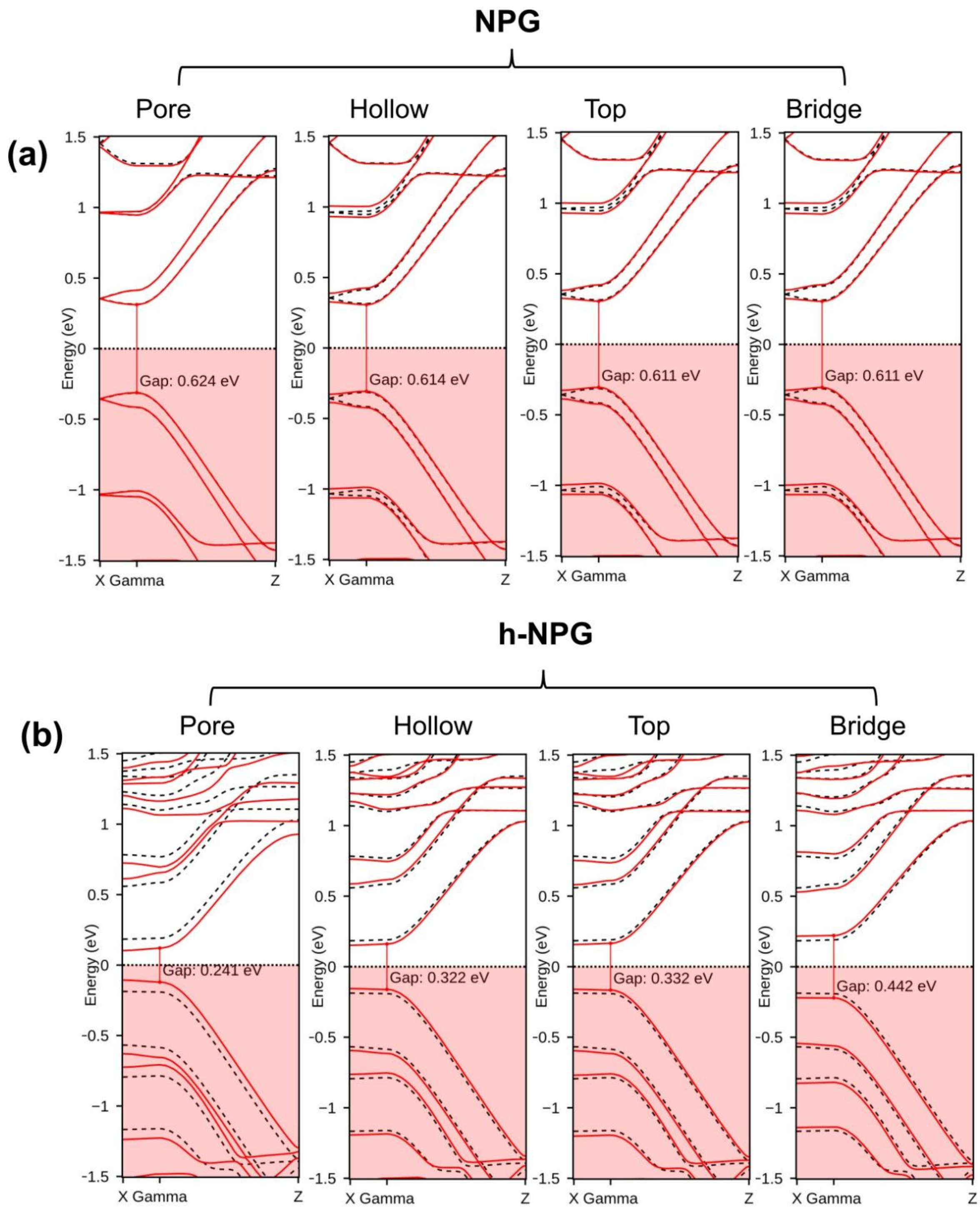

**Figure S2:** Band structures of **(a)** NPG and **(b)** h-NPG with water adsorbed at different sites: pore, hollow, top, and bridge. The black dotted lines in the band structures represent the corresponding pristine counterparts. The Fermi energy level is shifted to zero.

## 3. Band Structures of NPG/h-NPG with an Increasing Number of Water Molecules

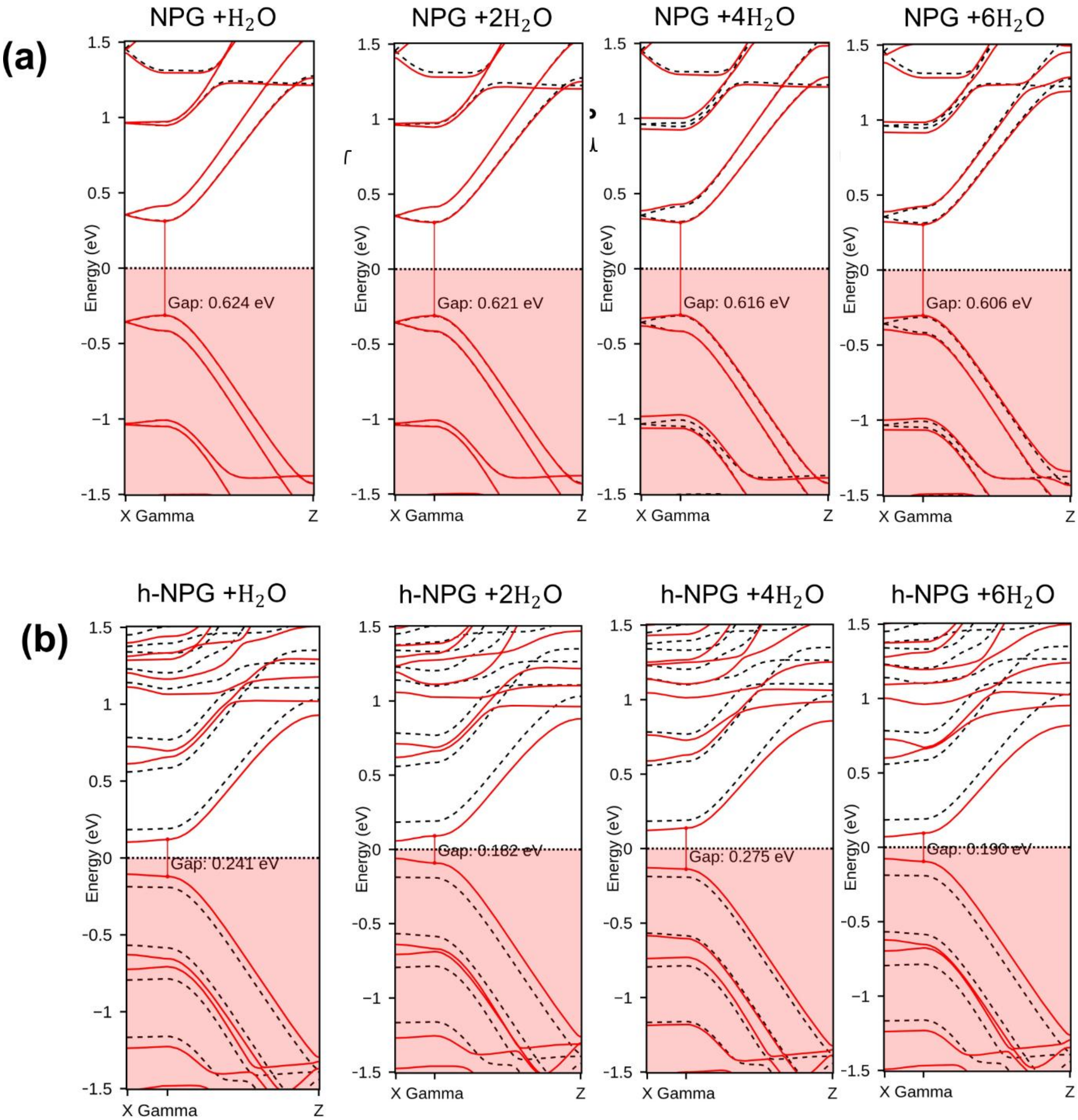

**Figure S3:** Band structures of (a) NPG and (b) h-NPG with an increasing number of water molecules (1, 2, and 3 water molecules in each pore). The black dotted lines in the band structures represent the corresponding pristine counterparts. The Fermi energy level is shifted to zero.

## 4. Electrostatic gating of Single-Water NPG System

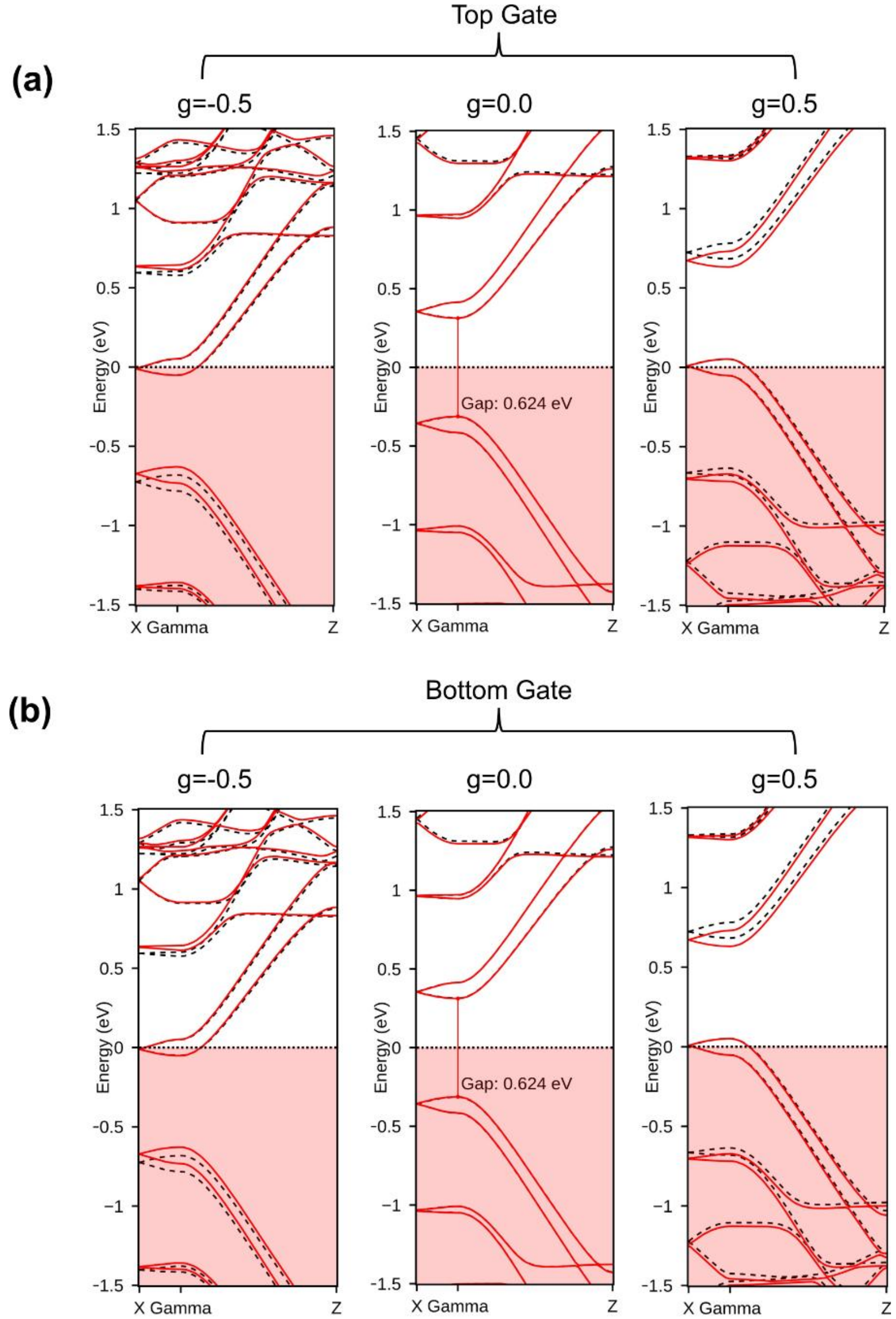

**Figure S4:** Band structure of single-water NPG under (a) top electrostatic gating and (b) bottom electrostatic gating. The number of injected electrons corresponds to $g \times 10^{13}\ e/cm^2$. The band filling reflects the occupation of the electronic states. For comparison, the band structures of pristine NPG under the same electrostatic gating conditions are shown as black dotted lines.

## 5. Electrostatic gating of Single-Water h-NPG System

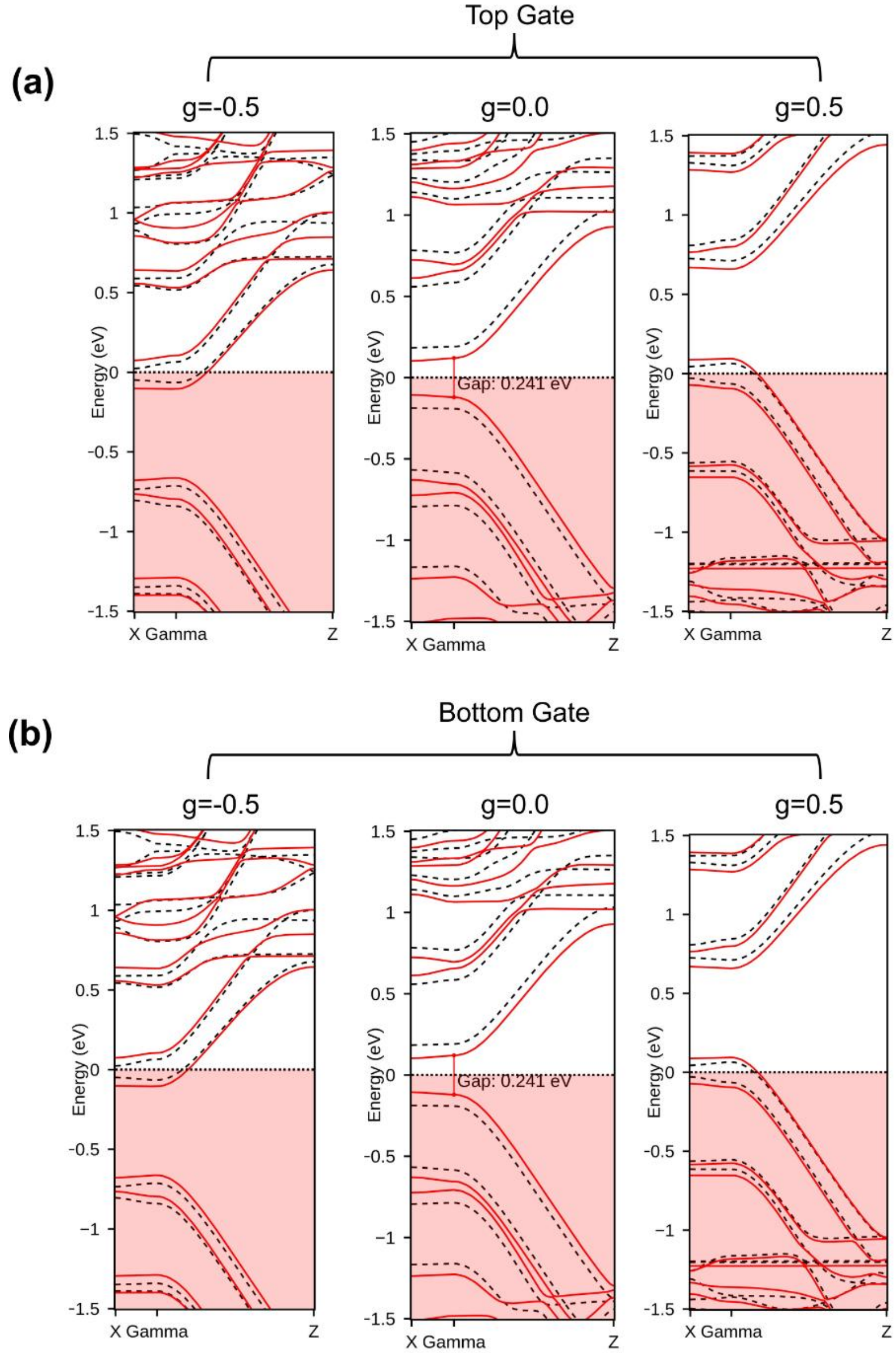

**Figure S5:** Band structure of single-water h-NPG under (a) top electrostatic gating and (b) bottom electrostatic gating. The number of injected electrons corresponds to $g \times 10^{13}\ e/cm^2$. The band filling reflects the occupation of the electronic states. For comparison, the band structures of pristine NPG under the same electrostatic gating conditions are shown as black dotted lines.

## 6. Bandgap Histograms of NPG/h-NPG under Single and Bulk-water Environments

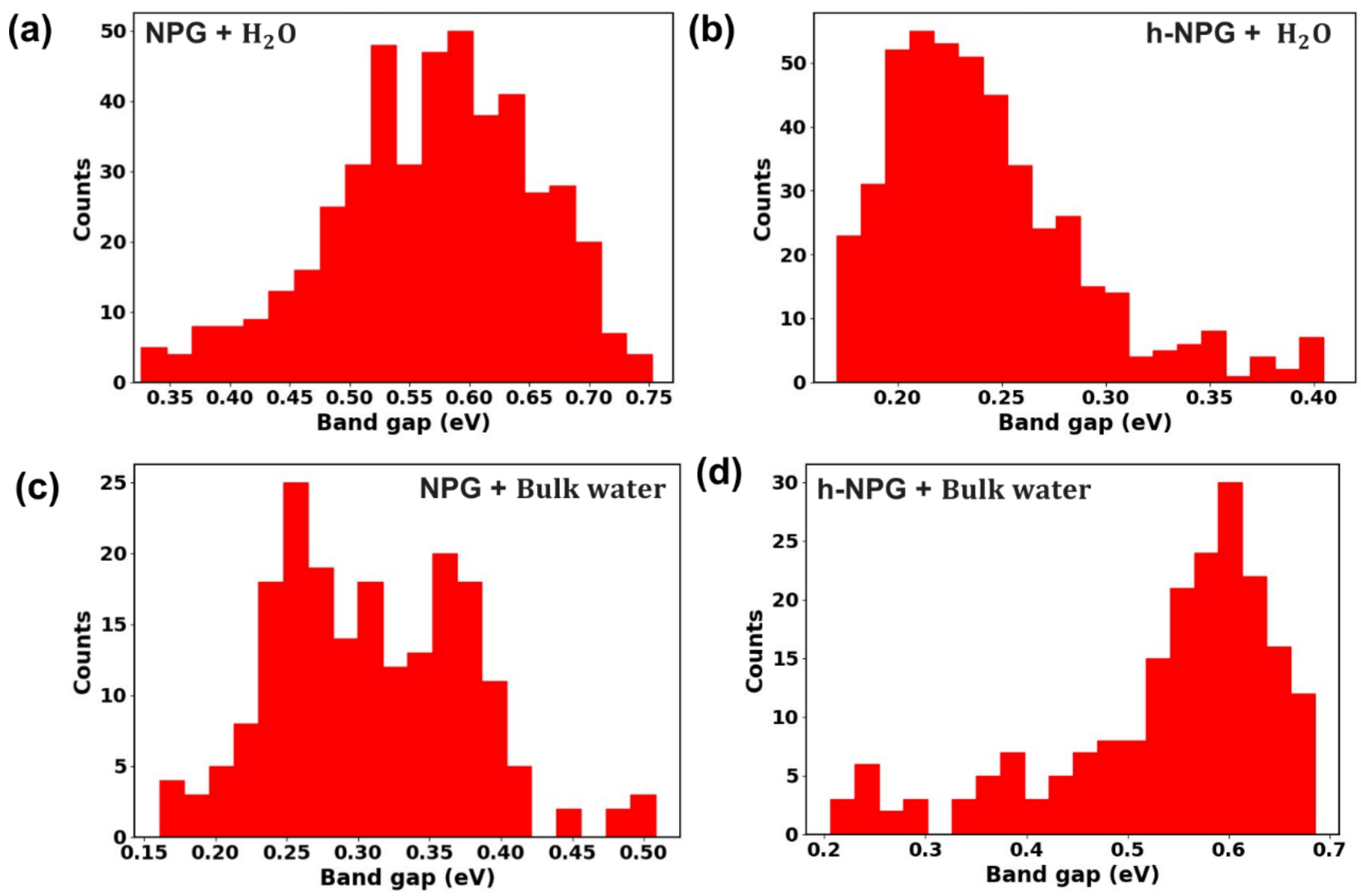

**Figure S6:** Bandgap histograms of (a) NPG and (b) h-NPG from AIMD-sampled single water configurations. Bandgap histograms of (c) NPG and (d) h-NPG from AIMD sampled bulk-water configurations.

**7. Tuned Hyperparameters and Model Performance**

**Table S1**: Optimized hyperparameters used for ML regression models for bandgap prediction of NPG and h-NPG systems under single and bulk water environments.

| Regression model | Optimized hyperparameters |
|---|---|
| Gaussian Process Regression (GPR) | Kernel: $C$ (1.0) × Matern (length scale = 1.0, $\nu$ = 1.5) + White Kernel; $\alpha = 1 \times 10^{-6}$; random state = 42 |
| Kernel Ridge Regression (KRR) | $\alpha$ = 0.001; polynomial kernel; degree = 2; $\gamma$ = 0.01; coef0 = 0.5 |
| Random Forest Regression (RFR) | Number of trees = 400; maximum depth = 20; minimum samples split = 2; random state = 42 |
| Extreme Gradient Boosting Regression (XGBR) | Number of trees = 300; maximum depth = 6; learning rate = 0.05; subsample = 0.8; column sample by tree = 0.9; objective = squared error; random state = 42 |

**Table S2**: Train and test RMSE values (in eV) of ML regression models for bandgap prediction of NPG/h-NPG systems. NPG + $H_2O$ and h-NPG + $H_2O$ denote single water systems, whereas NPG + BW and h-NPG + BW denote bulk-water systems.

| Regression model | RMSE | NPG + $H_2O$ | h-NPG + $H_2O$ | NPG + BW | h-NPG + BW |
|---|---|---|---|---|---|
| Gaussian Process Regression (GPR) | Train | 0.0004 | 0.0007 | 0.0007 | 0.0006 |
| | Test | 0.0217 | 0.0013 | 0.0041 | 0.0195 |
| Kernel Ridge Regression (KRR) | Train | 0.0006 | 0.0000 | 0.0006 | 0.0002 |
| | Test | 0.0238 | 0.0001 | 0.0019 | 0.0385 |
| Random Forest Regression (RFR) | Train | 0.0174 | 0.0052 | 0.0077 | 0.0217 |
| | Test | 0.0460 | 0.0139 | 0.0161 | 0.0740 |
| Extreme Gradient Boosting Regression (XGBR) | Train | 0.0006 | 0.0004 | 0.0005 | 0.0005 |
| | Test | 0.0427 | 0.0082 | 0.0129 | 0.0499 |

## 8. Confidence Intervals for Black-box GPR Models

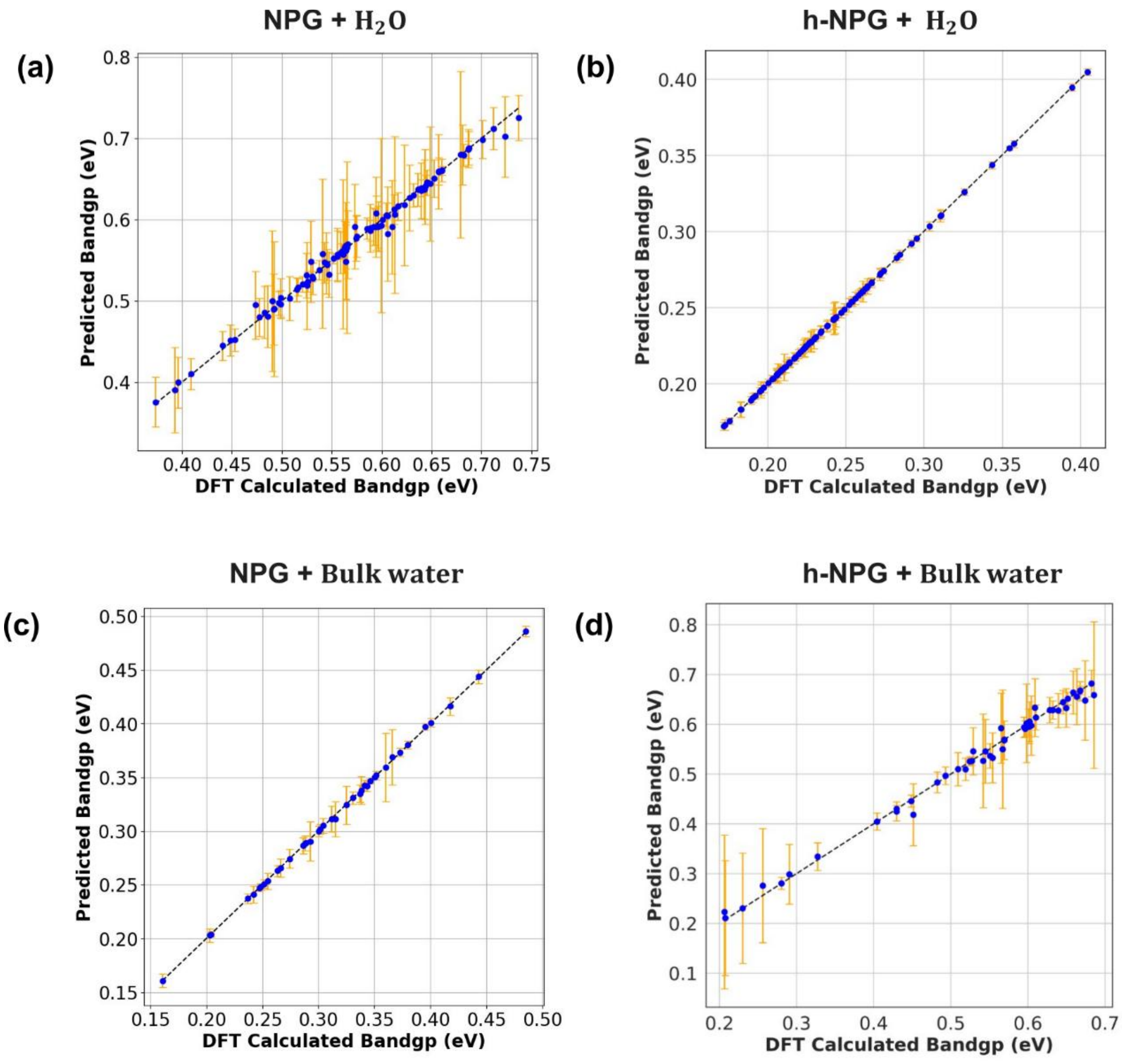

**Figure S7**: GPR-predicted versus DFT-calculated bandgaps using the PCA-reduced SOAP feature space. The plots show predictions for (a) NPG + $H_2O$, (b) h-NPG + $H_2O$, (c) NPG + Bulk water, and (d) h-NPG + Bulk water. The dashed diagonal line represents ideal agreement between predicted and DFT-calculated bandgaps, while the error bars indicate the 95% confidence intervals estimated from the predictive standard deviation of the GPR model.

## 9. Learning Curve Analysis for Black-box GPR Models

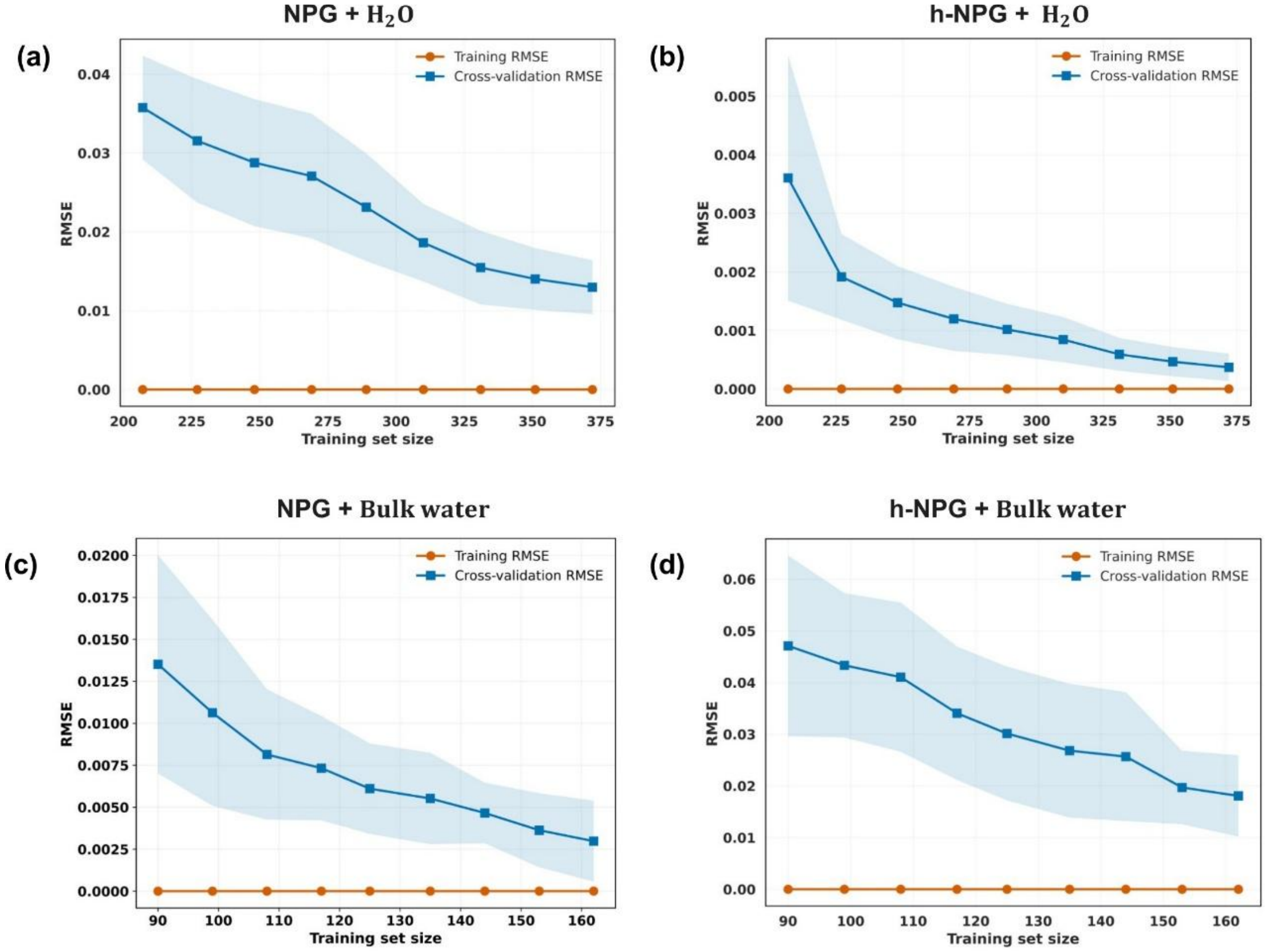

**Figure S8**: GPR-predicted versus DFT-calculated bandgaps using the PCA-reduced SOAP feature space. The plots show predictions for (a) NPG + $H_2O$, (b) h-NPG + $H_2O$, (c) NPG + Bulk water, and (d) h-NPG + Bulk water. The dashed diagonal line represents ideal agreement between predicted and DFT-calculated bandgaps, while the error bars indicate the 95% confidence intervals estimated from the predictive standard deviation of the GPR model.

**10. Supporting Note 1:**

**Generation for Grey-Box ML Bandgap Prediction in Single-Water NPG/h-NPG Systems**

We extracted a total of thirteen features for bandgap prediction in the single-water NPG/h-NPG systems. The extracted features and their formulations are described below.

**1. Center of Geometry (COG)**

The center of geometry of the water molecule is defined as the arithmetic mean of the positions of oxygen and hydrogen atoms:

$$\mathbf{R}_{\mathrm{COG}} = \frac{1}{3}\sum_{i=1}^{3}\mathbf{r}_i \qquad (1)$$

where $\mathbf{r}_i$ is the position vector of atom $i$ in the water molecule ($i = 1$ for O and $i = 2,3$ for H). The three Cartesian components are denoted $\mathrm{COG}_x(\mathrm{H_2O})$, $\mathrm{COG}_y(\mathrm{H_2O})$, and $\mathrm{COG}_z(\mathrm{H_2O})$, consistent with Table 1. They capture the spatial location of the water molecule within the NPG/h-NPG pore.

**2. Minimum N–H Distance ($\boldsymbol{d}_{\mathbf{min}}(\mathbf{H_w}, \mathbf{N})$)**

The minimum Euclidean distance between the hydrogen atoms of the water molecule and the nitrogen atoms of the h-NPG system is calculated as:

$$d_{\mathrm{min}}(\mathrm{H_w}, \mathrm{N}) = \min_{j,k} \left\| \mathbf{r}_{\mathrm{H}_{\mathrm{w},j}} - \mathbf{r}_{\mathrm{N}_k} \right\| \qquad (2)$$

where $\mathbf{r}_{\mathrm{H}_{\mathrm{w},j}}$ is the position of water hydrogen atom $j$ and $\mathbf{r}_{\mathrm{N}_k}$ is the position of N-dopant atom $k$ in h-NPG.

## 3. Minimum O–H Distance ($\boldsymbol{d_{\mathrm{min}}(\mathrm{O_w}, \mathrm{H_{pore}})}$)

The minimum Euclidean distance between the oxygen atom of the water molecule and the hydrogen-passivated pore edges of NPG/h-NPG is calculated as:

$$d_{\mathrm{min}}(\mathrm{O_w}, \mathrm{H_{pore}}) = \min_{m} \left\| \mathbf{r}_{\mathrm{O_w}} - \mathbf{r}_{\mathrm{H_{pore},\mathit{m}}} \right\| \qquad (3)$$

where $\mathbf{r}_{\mathrm{O_w}}$ is the position of the water oxygen and $\mathbf{r}_{\mathrm{H_{pore},\mathit{m}}}$ is the position of pore-edge hydrogen atom $m$.

## 4. Dipole Moment of Water

The molecular dipole of the water molecule is computed relative to its center of geometry:

$$\boldsymbol{\mu} = \sum_{i=1}^{3} q_i \, (\mathbf{r}_i - \mathbf{R}_{\mathrm{COG}}) \qquad (4)$$

where $q_i$ is the partial charge of atom $i$ (oxygen: $-2\,|e|$; hydrogen: $+1\,|e|$), $\mathbf{r}_i$ is its position vector, and $\mathbf{R}_{\mathrm{COG}}$ is the centre of geometry. The Cartesian components are labelled $\mu_x(\mathrm{H_2O})$, $\mu_y(\mathrm{H_2O})$, and $\mu_z(\mathrm{H_2O})$ in Table 1. The resulting dipole vector, expressed in $\mathrm{e \cdot \AA}$, is converted to Debye using the factor $4.803\ \mathrm{D}/(\mathrm{e \cdot \AA})$. Its magnitude is:

$$|\boldsymbol{\mu}(\mathrm{H_2O})| = \sqrt{\mu_x^2(\mathrm{H_2O}) + \mu_y^2(\mathrm{H_2O}) + \mu_z^2(\mathrm{H_2O})} \qquad (5)$$

where $\mu_x$, $\mu_y$, and $\mu_z$ are the Cartesian components of the dipole vector. Together, the dipole components and magnitude describe the orientation and polarity of the water molecule.

## 5. Dipole Angle ($\boldsymbol{\theta_{\mu,z}}$)

The orientation of the water dipole moment relative to the surface normal is described by the angle between the water dipole vector and the $z$-axis:

$$\theta_{\mu,z} = \cos^{-1}\left(\frac{\boldsymbol{\mu} \cdot \hat{\mathbf{z}}}{|\boldsymbol{\mu}|}\right)\frac{180}{\pi} \qquad (6)$$

where $\hat{\mathbf{z}}$ is the unit vector along the surface-normal direction. The resulting value is reported in degrees.

## 6. Water–Substrate Distance Descriptors

To describe the position of the water molecule relative to the graphene carbon framework, the Euclidean distance between the water COG and each carbon atom is calculated as:

$$d(\mathrm{H_2O}, \mathrm{C}_k) = \|\mathbf{R}_{\mathrm{COG}} - \mathbf{r}_{\mathrm{C}_k}\|, \quad k = 1,2, \dots, N_{\mathrm{C}} \qquad (7)$$

where $\mathbf{r}_{\mathrm{C}_k}$ is the position of carbon atom $k$ and $N_{\mathrm{C}}$ is the total number of carbon atoms in the NPG/h-NPG framework. From the distance set $\{d(\mathrm{H_2O}, \mathrm{C}_k)\}_{k=1}^{N_{\mathrm{C}}}$, three descriptors are obtained:

$$d_{\min}(\mathrm{H_2O}, \mathrm{C}) = \min_k d(\mathrm{H_2O}, \mathrm{C}_k) \qquad (8)$$

$$d_{\max}(\mathrm{H_2O}, \mathrm{C}) = \max_k d(\mathrm{H_2O}, \mathrm{C}_k) \qquad (9)$$

$$\sigma_d[d(\mathrm{H_2O}, \mathrm{C})] = \sqrt{\frac{1}{N_{\mathrm{C}}}\sum_{k=1}^{N_{\mathrm{C}}}\left(d(\mathrm{H_2O}, \mathrm{C}_k) - \overline{d(\mathrm{H_2O}, \mathrm{C})}\right)^2} \qquad (10)$$

where $\overline{d(\mathrm{H_2O, C})}$ is the mean distance between the water COG and framework carbon atoms.

## 11. Supporting Note 2:

### Feature Generation for Bulk-Water NPG/h-NPG Systems

A total of 38 features were extracted to characterize the structural and electrostatic environment of bulk-water NPG/h-NPG systems. These features were computed from the atomic coordinates, with particular emphasis on water molecules and their interactions with the substrate. Distances and the selection of local dipole environments were evaluated using a cutoff radius of $r_\mathrm{c} = 5$ Å to retain physically relevant interactions. The feature set combines global descriptors of the complete water layer with spatially resolved descriptors obtained by partitioning the confined water into distinct regions.

### i. Centre of Geometry of Water

For water molecule $i$, the centre of geometry (COG) is the arithmetic mean of the positions of its oxygen and two hydrogen atoms:

$$\mathbf{R}_{\mathrm{COG},i} = \frac{1}{3}\sum_{a=1}^{3} \mathbf{r}_{i,a} \qquad (1)$$

where $\mathbf{r}_{i,a}$ is the position vector of atom $a$ in water molecule $i$. For each Cartesian component $\alpha \in \{x, y, z\}$, the mean COG coordinates across the $N_\mathrm{w}$ water molecules is:

$$\mathrm{mean}[\mathrm{COG}_\alpha(\mathrm{H_2O})] = \frac{1}{N_\mathrm{w}}\sum_{i=1}^{N_\mathrm{w}} R_{\mathrm{COG},i,\alpha} \qquad (2)$$

and its standard deviation is:

$$\sigma_d[\mathrm{COG}_\alpha(\mathrm{H_2O})] = \sqrt{\frac{1}{N_\mathrm{w}}\sum_{i=1}^{N_\mathrm{w}}\left(R_{\mathrm{COG},i,\alpha} - \mathrm{mean}[\mathrm{COG}_\alpha(\mathrm{H_2O})]\right)^2} \qquad (3)$$

The six descriptors are $\mathrm{mean}\left[\mathrm{COG}_{x/y/z}(\mathrm{H_2O})\right]$ and $\sigma_d\left[\mathrm{COG}_{x/y/z}(\mathrm{H_2O})\right]$, consistent with Table 2. The mean coordinates describe the average spatial distribution of water, whereas the standard deviations quantify its spatial dispersion.

**ii. Global Water Dipole-Moment Features**

The dipole moment of water molecule $i$, evaluated relative to its COG, is:

$$\boldsymbol{\mu}_i = \sum_{a=1}^{3} q_{i,a}\left(\mathbf{r}_{i,a} - \mathbf{R}_{\mathrm{COG},i}\right) \qquad (4)$$

where $q_{i,a}$ and $\mathbf{r}_{i,a}$ are the charge and position vector of atom $a$ in molecule $i$. The total dipole vector of the water layer is:

$$\boldsymbol{\mu}^{\mathrm{tot}} = \sum_{i=1}^{N_\mathrm{w}} \boldsymbol{\mu}_i \qquad (5)$$

with Cartesian components $\mu_x^{\mathrm{tot}}$, $\mu_y^{\mathrm{tot}}$, and $\mu_z^{\mathrm{tot}}$. Its magnitude is:

$$|\boldsymbol{\mu}^{\mathrm{tot}}| = \sqrt{(\mu_x^{\mathrm{tot}})^2 + \left(\mu_y^{\mathrm{tot}}\right)^2 + (\mu_z^{\mathrm{tot}})^2} \qquad (6)$$

The mean and standard deviation of the individual molecular dipole magnitudes are:

$$\mathrm{mean}[|\boldsymbol{\mu}|] = \frac{1}{N_\mathrm{w}}\sum_{i=1}^{N_\mathrm{w}}|\boldsymbol{\mu}_i| \qquad (7)$$

$$\sigma_d[|\boldsymbol{\mu}|] = \sqrt{\frac{1}{N_\mathrm{w}}\sum_{i=1}^{N_\mathrm{w}}(|\boldsymbol{\mu}_i| - \mathrm{mean}[|\boldsymbol{\mu}|])^2} \qquad (8)$$

The six global dipole descriptors are $\mu_{x/y/z}^{\mathrm{tot}}$, $|\boldsymbol{\mu}^{\mathrm{tot}}|$, $\mathrm{mean}[|\boldsymbol{\mu}|]$, and $\sigma_d[|\boldsymbol{\mu}|]$, matching Table 2. They describe collective polarization, average molecular polarization strength, and polarization heterogeneity.

**iii. Dipole Angle Relative to the Surface**

For each water molecule, the angle between its dipole vector and the surface-normal direction $\hat{\mathbf{z}}$ is:

$$\theta_{\mu,z,i} = \cos^{-1}\left(\frac{\boldsymbol{\mu}_i \cdot \hat{\mathbf{z}}}{|\boldsymbol{\mu}_i|}\right)\frac{180}{\pi} \qquad (9)$$

The mean dipole angle and its standard deviation are:

$$\mathrm{mean}[\theta_{\mu,z}] = \frac{1}{N_\mathrm{w}}\sum_{i=1}^{N_\mathrm{w}}\theta_{\mu,z,i} \qquad (10)$$

$$\sigma_d[\theta_{\mu,z}] = \sqrt{\frac{1}{N_\mathrm{w}}\sum_{i=1}^{N_\mathrm{w}}(\theta_{\mu,z,i} - \mathrm{mean}[\theta_{\mu,z}])^2} \qquad (11)$$

These descriptors quantify the average orientation and orientational disorder of the water dipoles relative to the substrate.

### iv. Intermolecular O···H Distances

The intermolecular O···H distance set contains distances between oxygen atom $\mathrm{O}_i$ of water molecule $i$ and hydrogen atom $\mathrm{H}_{j,a}$ of a different water molecule $j$:

$$\mathcal{D}_{\mathrm{OH}} = \left\{ \left\| \mathbf{r}_{\mathrm{O}_i} - \mathbf{r}_{\mathrm{H}_{j,a}} \right\| : \; i \neq j, \; d_{\mathrm{OH}} \leq r_{\mathrm{c}} \right\} \qquad (12)$$

Intramolecular O-H pairs are excluded. The minimum and mean distances are:

$$\min[d(\mathrm{O}_{\mathrm{w}}, \mathrm{H}_{\mathrm{w}})] = \min_{d \in \mathcal{D}_{\mathrm{OH}}} d \qquad (13)$$

$$\mathrm{mean}[d(\mathrm{O}_{\mathrm{w}}, \mathrm{H}_{\mathrm{w}})] = \frac{1}{|\mathcal{D}_{\mathrm{OH}}|} \sum_{d \in \mathcal{D}_{\mathrm{OH}}} d \qquad (14)$$

Short O···H distances indicate strong hydrogen bonding within the water network, which enhances local polarization and may indirectly influence the substrate electronic structure.

### v. Water-Substrate N···H Distances

The N···H distance set contains distances between water hydrogen atoms and nitrogen atoms of the h-NPG substrate within the cutoff:

$$\mathcal{D}_{\mathrm{NH}} = \left\{ \left\| \mathbf{r}_{\mathrm{N}_k} - \mathbf{r}_{\mathrm{H}_{i,a}} \right\| : \; d_{\mathrm{NH}} \leq r_{\mathrm{c}} \right\} \qquad (15)$$

The minimum and mean N···H distances are:

$$\min[d(\mathrm{N}, \mathrm{H}_{\mathrm{w}})] = \min_{d \in \mathcal{D}_{\mathrm{NH}}} d \qquad (16)$$

$$\mathrm{mean}[d(\mathrm{N}, \mathrm{H}_{\mathrm{w}})] = \frac{1}{|\mathcal{D}_{\mathrm{NH}}|} \sum_{d \in \mathcal{D}_{\mathrm{NH}}} d \qquad (17)$$

Short N···H distances indicate strong donor–acceptor interactions that can perturb the local electronic environment and shift the conduction- or valence-band edges.

**vi. Number of Contacts**

The numbers of O···H and N···H contacts within the cutoff are defined as:

$$N_{\mathrm{OH}} = \sum_{i \neq j} \sum_{a} \mathbb{I}\left(d_{\mathrm{O}_i \mathrm{H}_{j,a}} \leq r_{\mathrm{c}}\right) \qquad (18)$$

$$N_{\mathrm{NH}} = \sum_{k} \sum_{i,a} \mathbb{I}\left(d_{\mathrm{N}_k \mathrm{H}_{i,a}} \leq r_{\mathrm{c}}\right) \qquad (19)$$

where $\mathbb{I}(\cdot)$ is 1 when the stated condition is satisfied and 0 otherwise. These contact counts provide global measures of the water hydrogen-bond network density and water–substrate electrostatic coupling.

**vii. Region-Resolved Dipole Features**

The confined water layer is partitioned along the pore direction into a bridge region ($B$), a left ribbon, and a right ribbon. The bridge is the inter-pore channel between the two inner pore-edge boundaries. The outer hydration regions form the left and right ribbons; because they are connected through periodic boundary conditions, they are combined into a single left/right ribbon region ($LR$).

For a region $R \in \{\text{bridge}, \text{left/right}\}$ containing $N_R$ water molecules, the region-resolved total dipole vector is:

$$\boldsymbol{\mu}^{R} = \sum_{i \in R} \boldsymbol{\mu}_i \qquad (20)$$

and its magnitude is:

$$|\boldsymbol{\mu}^R| = \sqrt{(\mu_x^R)^2 + \left(\mu_y^R\right)^2 + (\mu_z^R)^2} \qquad (21)$$

The regional mean and standard deviation of individual dipole magnitudes are:

$$\mathrm{mean}[|\boldsymbol{\mu}^R|] = \frac{1}{N_R}\sum_{i\in R}|\boldsymbol{\mu}_i| \qquad (22)$$

$$\sigma_d[|\boldsymbol{\mu}^R|] = \sqrt{\frac{1}{N_R}\sum_{i\in R}(|\boldsymbol{\mu}_i| - \mathrm{mean}[|\boldsymbol{\mu}^R|])^2} \qquad (23)$$

For the bridge region, the six descriptors are $\mu_{x/y/z}^{\mathrm{bridge}}$, $\left|\boldsymbol{\mu}^{\mathrm{bridge}}\right|$, $\mathrm{mean}\left[\left|\boldsymbol{\mu}^{\mathrm{bridge}}\right|\right]$, and $\sigma_d\left[\left|\boldsymbol{\mu}^{\mathrm{bridge}}\right|\right]$. The corresponding left/right descriptors are $\mu_{x/y/z}^{\mathrm{left/right}}$, $\left|\boldsymbol{\mu}^{\mathrm{left/right}}\right|$, $\mathrm{mean}\left[\left|\boldsymbol{\mu}^{\mathrm{left/right}}\right|\right]$, and $\sigma_d\left[\left|\boldsymbol{\mu}^{\mathrm{left/right}}\right|\right]$.

**viii. Bridge-Ribbon Difference Features:**

For any regional descriptor $X$, the bridge-left/right difference is defined as:

$$\Delta X^{\mathrm{bridge-left/right}} = X^{\mathrm{bridge}} - X^{\mathrm{left/right}} \qquad (24)$$

Accordingly, the six polarization-contrast descriptors are:

$$\Delta\mu_\alpha = \mu_\alpha^{\mathrm{bridge}} - \mu_\alpha^{\mathrm{left/right}}, \quad \alpha \in \{x, y, z\} \qquad (25)$$

$$\Delta|\boldsymbol{\mu}|^{\mathrm{bridge-left/right}} = \left|\boldsymbol{\mu}^{\mathrm{bridge}}\right| - \left|\boldsymbol{\mu}^{\mathrm{left/right}}\right| \qquad (26)$$

$$\Delta\mathrm{mean}[|\boldsymbol{\mu}|]^{\mathrm{bridge-left/right}} = \mathrm{mean}\left[\left|\boldsymbol{\mu}^{\mathrm{bridge}}\right|\right] - \mathrm{mean}\left[\left|\boldsymbol{\mu}^{\mathrm{left/right}}\right|\right] \qquad (27)$$

$$\Delta\sigma_d[|\boldsymbol{\mu}|]^{\text{bridge}-\text{left/right}} = \sigma_d\left[\left|\boldsymbol{\mu}^{\text{bridge}}\right|\right] - \sigma_d\left[\left|\boldsymbol{\mu}^{\text{left/right}}\right|\right] \qquad (28)$$

Together, these features quantify the polarization contrast between water molecules in the bridge region and those in the combined left/right ribbon region.

## 12. Confidence Intervals for Grey-box GPR Models

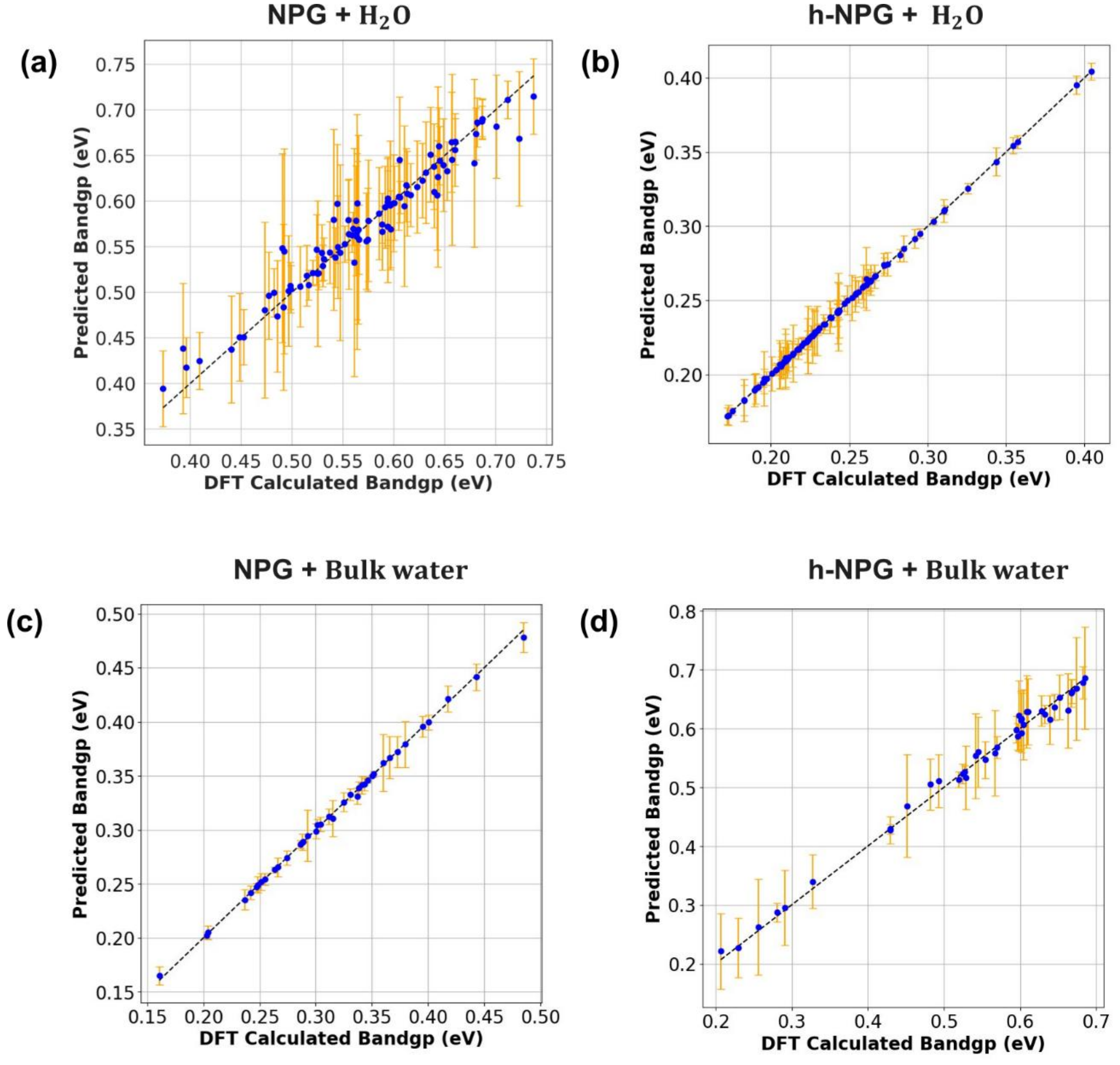

**Figure S9**: GPR-predicted versus DFT-calculated bandgaps using the physics-informed descriptors. The plots show predictions for (a) NPG + $H_2O$, (b) h-NPG + $H_2O$, (c) NPG + Bulk water, and (d) h-NPG + Bulk water. The dashed diagonal line represents ideal agreement between predicted and DFT-calculated bandgaps, while the error bars indicate the 95% confidence intervals estimated from the predictive standard deviation of the GPR model.

## 13. Learning Curve Analysis for Grey-box GPR Models

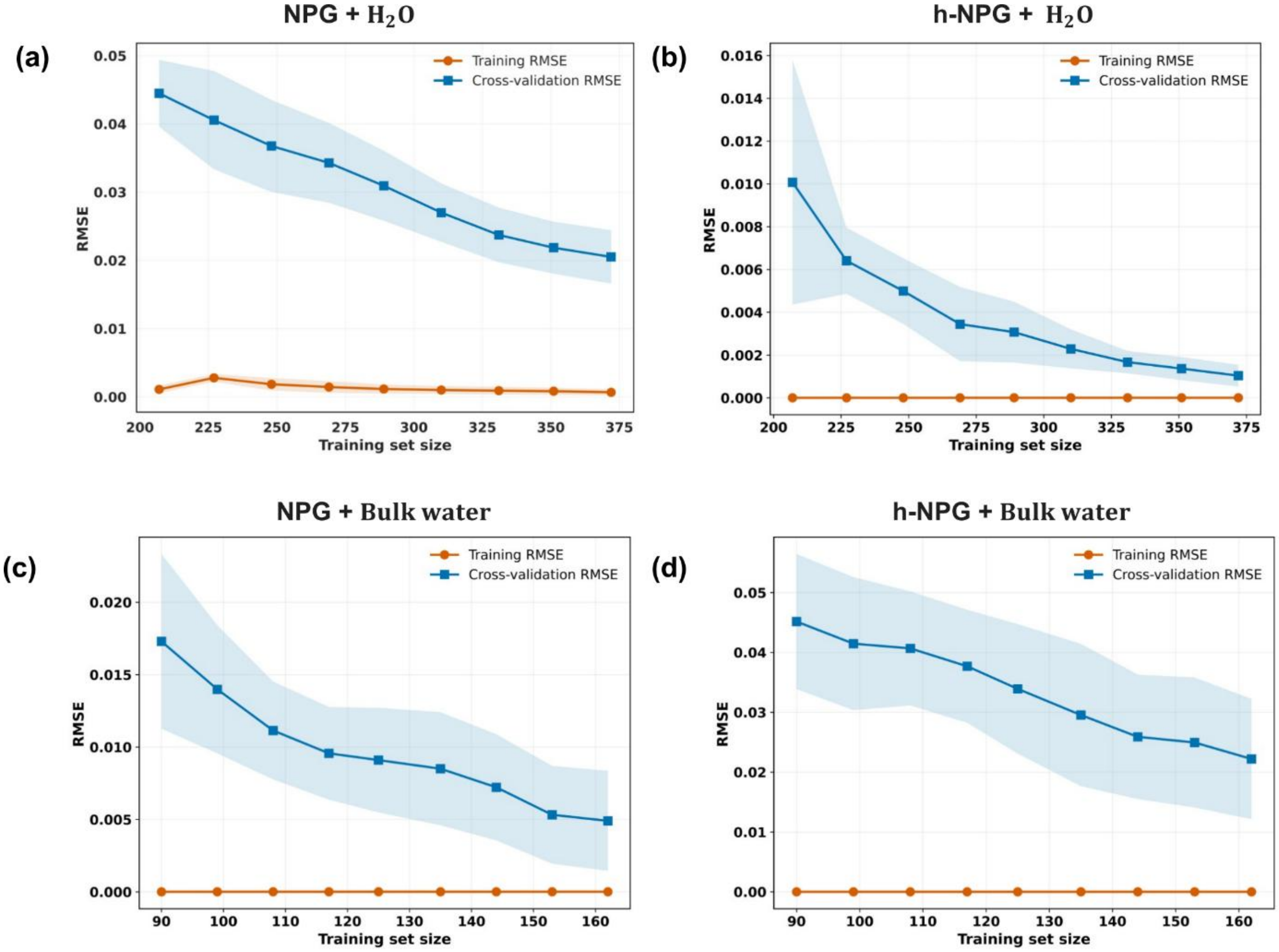

**Figure S10**: GPR-predicted versus DFT-calculated bandgaps using the physics-informed descriptors. The plots show predictions for (a) NPG + $H_2O$, (b) h-NPG + $H_2O$, (c) NPG + Bulk water, and (d) h-NPG + Bulk water. The dashed diagonal line represents ideal agreement between predicted and DFT-calculated bandgaps, while the error bars indicate the 95% confidence intervals estimated from the predictive standard deviation of the GPR model.